\documentclass[]{article}

\usepackage{amsmath, amsthm, amssymb, amsfonts}
\usepackage{mathtools}
\usepackage{hyperref}
\usepackage{graphicx}
\usepackage{dsfont}
\usepackage{fullpage}
\usepackage{color}
\usepackage{soul} 
\usepackage[title]{appendix}
\usepackage{tikz}
\usetikzlibrary{patterns}
\usetikzlibrary{shapes.multipart}
\usetikzlibrary{arrows}

\theoremstyle{definition}

\definecolor{labelkey}{cmyk}{.4,.2,0,0}

\newcommand{\be}{\begin{equation}}
\newcommand{\ee}{\end{equation}}
\newcommand{\bea}{\begin{eqnarray}}
\newcommand{\eea}{\end{eqnarray}}

\usepackage{titlesec}
\titleformat{\section}{\large\bf}{\thesection}{1em}{}
\titleformat{\subsection}[runin]{\bf}{\thesubsection}{1em}{}[.]

\usepackage{authblk}
\title{This is some thing}
\author[1]{Pierre Le Doussal}
\affil[1]{\normalsize Laboratoire de Physique de l'\'Ecole Normale Sup\'erieure, ENS, Universit\'e PSL, CNRS, Sorbonne Universit\'e, Universit\'e de Paris, 75005 Paris, France}

\title{\bf \large Dynamic scaling of growing interfaces }
\date{}

\begin{document}

\maketitle

\begin{abstract}
We give a brief overview of the seminal paper which introduced the Kardar-Parisi-Zhang equation as a paradigmatic model
for random growth in 1986. We describe some of the developments to which it gave rise in mathematics and physics over the years, and some
examples of applications. 
\end{abstract}

\bigskip

\begin{center}
Contribution to the volume \\
{\it From Quantum Fields to Spin Glasses: \\
A journey through the contributions of Giorgio Parisi to theoretical Physics}
\end{center}


\section{Introduction}

It's a great pleasure to celebrate the famous 1986 KPZ paper \cite{KPZ} on interface growth. On a personal note, I became aware of the beauty of this work during my postdoc in 1989 when
I met Mehran Kardar and David Nelson. At that time there was an explosion of activity about directed polymers and the KPZ equation, with great excitement about applications to vortex lines in high-$T_c$ superconductors \cite{LarkinReview}. These years are often called the first KPZ revolution.
Since that time the KPZ equation has been with us, periodically reappearing in various reincarnations in physics and in mathematics. What is amazing about this work is that it established a bridge between growth and non-equilibrium physics (after all the KPZ equation is often called the Ising model of the non-equilibrium physics), and a host of other problems, such as the statistical mechanics of disordered systems. It also established a bridge between physics and mathematics which was amazingly fruitful in recent years. Finally, applications have been plethoric, and I will mention here
only on a small fraction of them. 

To gain perspective, let me mention the other important universality class for random growth in two dimensions, which is diffusion limited aggregation (DLA), introduced in another 
seminal paper \cite{Witten} (with about the same number of citations as of today, 7K). There Brownian particles 
come from infinity and stick to the cluster when they reach it. It is a very non-local rule, which leads to dendritic branching.
By contrast the KPZ growth describes generic local stochastic rules, and leads to much more gentle self-affine interfaces. Interestingly there are interpolations between these two types of growth.
These problems are difficult in euclidean space, but recently there was mathematical progress in fluctuating geometries, under the name of quantum Loewner evolution QLE$(\gamma^2,\eta)$
\cite{QLE}. It describes a two-parameter family of conformal growth of domains in the plane, and is related to Liouville quantum gravity (LQG)
(i.e. a random measure $d^2 x e^{\gamma \phi(x)}$ based on the Gaussian free field $\phi$). The growth rate per unit length is the
harmonic measure raised to the power $\eta$, so for $\eta=1$ it is like DLA but on a fluctuating geometry, and for $\eta=0$ it describes e.g. the Eden model on a random triangulation (i.e. an analog of KPZ growth). Hence in this field KPZ \cite{KPZ} 
naturally meets the other KPZ \cite{KPZ0,KPZPolyakov,KPZProof} ! (see \cite{footnoteKPZmeetKPZ}). 
\\

{\bf Outline}. I will start with the genesis of the KPZ equation, its connections with directed polymers and the Burgers equation, and why in one space dimension $d=1$ one can find the exact scaling exponents. Then briefly mention the situation in higher dimension $d >1$, which is mostly open questions. Then I will describe what is often called the second KPZ revolution, which happened
both in mathematics and in physics around 2000: in $d=1$ exact solutions of several discrete models were found to exhibit the same universal behavior at large scale (time and space), called the 1d KPZ class. It became possible to compute the full probability density function (PDF) of the height field at large time, unveiling amazing connections to universal distributions in 
random matrix theory (RMT), the so-called Tracy-Widom distributions. Then I will return to the KPZ equation itself, 
and its solution using replica. Studying the KPZ equation with replica started around 1990 with works of Kardar, and of Giorgio, but the full solution was only obtained after 2010. Non rigorously by physicists, but soon and in most cases, probabilists made them rigorous, extended them and unveiled in the process new mathematical structures leading to a huge pile of new solvable models. These developments are sometimes called a third KPZ revolution.
Mathematicians also proved non-trivial reassuring things such that (i) one can actually make mathematical sense of the KPZ equation (ii) the KPZ equation belongs to its own universality class. 
Finally, I will mention some interesting properties of two-time KPZ and close with 
a few chosen applications. 

\section{The KPZ paper and follow up} 

\begin{figure}[t]
     \includegraphics[width=0.5\columnwidth]{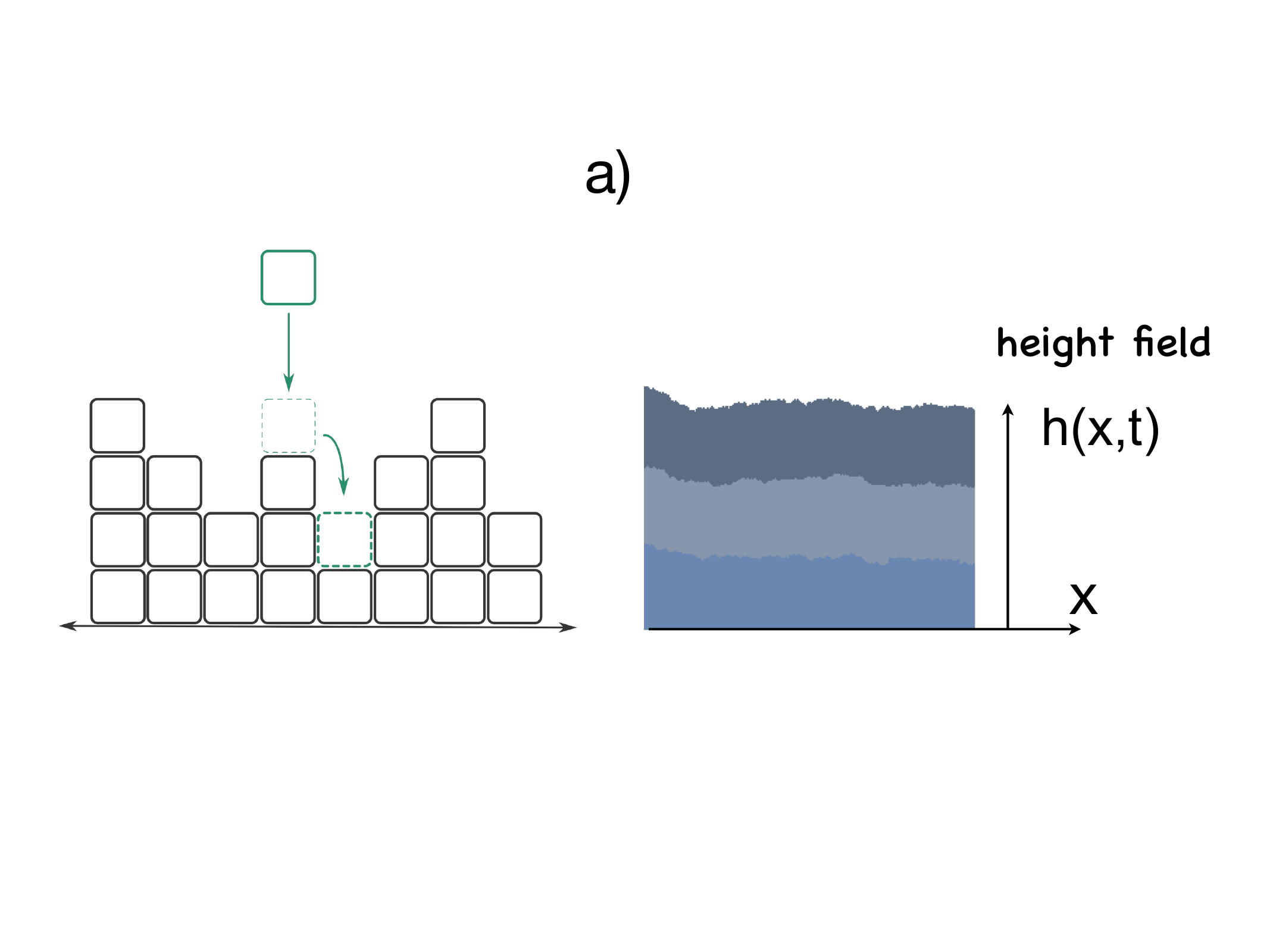}
      \includegraphics[width=0.5\columnwidth]{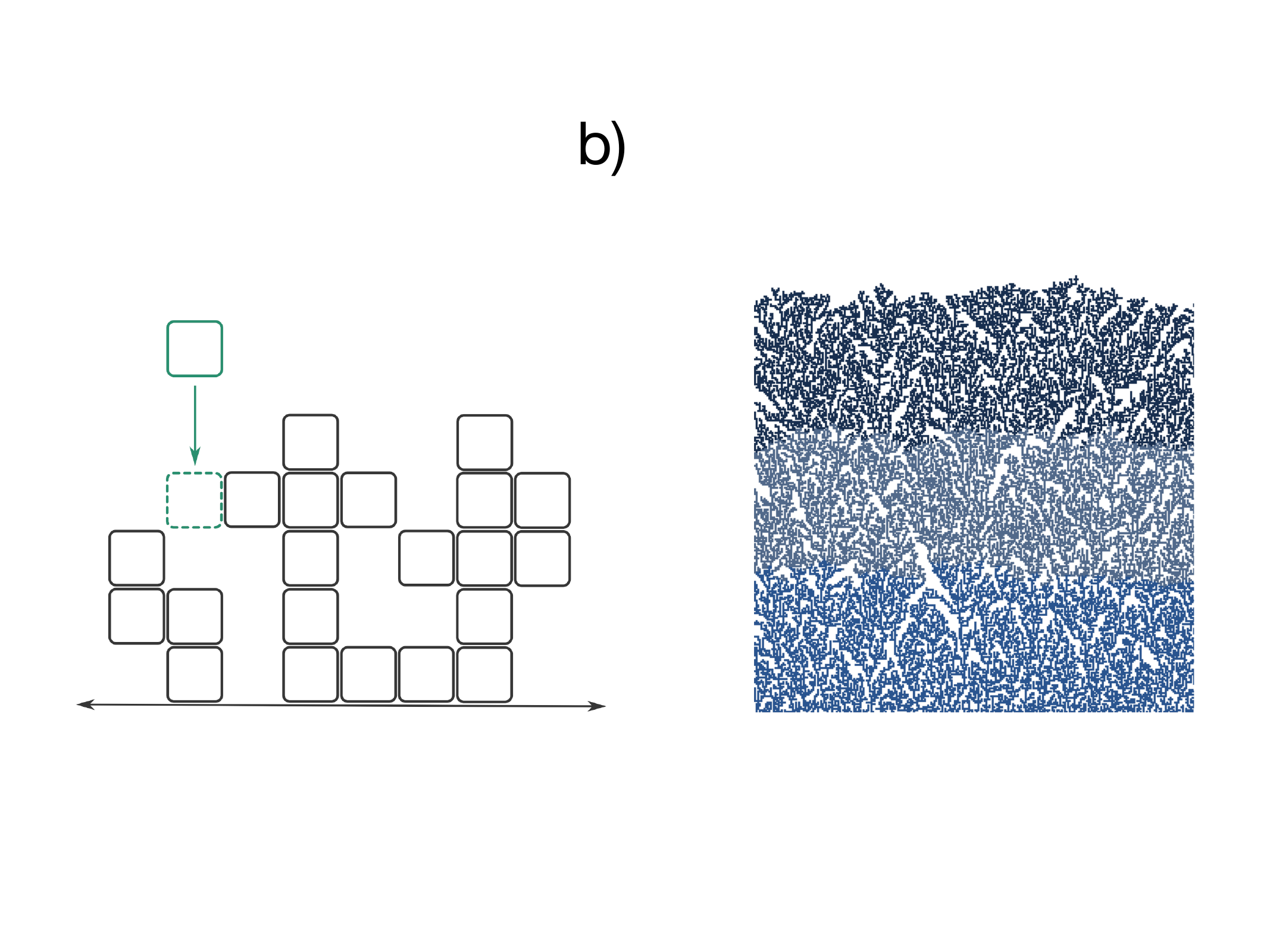}
     \vspace{-1.6cm}
   \caption{a) random deposition plus relaxation b) ballistic deposition} 
 \label{fig:Eden}
\end{figure}

The question asked in \cite{KPZ} is whether there is a universal description of the growth of an interface. Universality here means that, as for critical phenomena, one is looking for a coarse-grained model, based on symmetry considerations, which reproduces the large scale behavior.
Consider first $d=1$ and grains falling randomly on some surface. The height field $h(x,t)$ grows on average linearly in time,
and we are interested in its fluctuations. A coarse grained model is just linear growth plus noise and if there is a redistribution of grains locally, one adds a diffusive term, see Fig. \ref{fig:Eden} a) . This is the Edwards-Wilkinson (EW) equation for growth \cite{EW}
\be \label{EW} 
\partial_t h = \nu \partial_x^2 h + \eta(x,t) + v 
\ee
where $v+\eta(x,t)$ is the local growth rate, and if the growth rule is local,
$\eta(x,t)$ can be chosen as
space time white noise i.e. from a centered Gaussian distribution with covariance $\overline{\eta(x,t) \eta(x',t')}= 2 D \delta^d(x-x') \delta(t-t')$. Since it is a linear equation
it can be solved explicitly for any initial condition, and the height field has a Gaussian
distribution. Some details of the space-time height covariance depend on the initial condition but there is an overall scaling behavior of the height fluctuations $\delta h(x,t)=h(x,t)-v t$ 
\be 
\delta h \sim x^\chi \quad , \quad \delta h \sim t^\beta \quad , \quad z=\alpha/\beta 
\ee 
where $\chi$ (also called $\alpha$), $\beta$ and $z$ are respectively the roughness, growth and dynamical exponents.
For the EW model one finds $\chi_{EW}=1/2$, $\beta_{EW}=1/4$ and $z_{EW}=2$. It is a very
simple model, with diffusive dynamics, and in fact it describes an equilibrium situation. Indeed \eqref{EW} can be written in a detailed balance form (in the comoving frame, i.e. performing the shift $h \to h + v t$) 
\be 
\partial_t h = - \nu \frac{\delta H}{\delta h(x,t)}  + \eta(x,t) \quad , \quad H= \frac{1}{2} \int dx \,  (\partial_x h)^2 
\ee 
so that the height distribution converges at large time to the Gibbs equilibrium ${\cal P}_{eq}[h]
\sim e^{- \frac{\nu}{D} H[h]}$. In an infinite system this convergence holds for local correlations.
Hence the EW equation actually describes equilibrium growth: for instance, it is a coarse grained
model for a single domain wall performing Glauber dynamics in the ferromagnetic phase of the 2D Ising model without an external field
\cite{SpohnHouches2016}. It is thus too simple to capture irreversibility in the growth. 
For instance models of ballistic deposition where the grain can actually stick,
and example being the Eden model where all perimeter sites are equally likely
\cite{FamilyVicsekEdenBallistic85,Eden2}.
In these models, there is an additional phenomenon, which is lateral growth, see Fig. \ref{fig:Eden} b).
In \cite{KPZ}
they consider an interface which grows with some fixed velocity $v$ along its local normal direction.
If the interface is tilted locally by angle $\theta$ from horizontal, projecting 
along the vertical direction gives $\partial_t h=v/\cos \theta=v \sqrt{1+ (\partial_x h)^2}
\simeq v + \frac{v}{2} (\partial_x h)^2$ leading to the KPZ growth equation (written now in $d$ space
dimensions, $x \in \mathbb{R}^d$, for growth in $d+1$ dimension) 
\be \label{KPZ} 
\partial_t h = \nu \nabla^2 h + \frac{\lambda}{2} (\nabla h)^2 + \eta(x,t) 
\ee
again after the shift $h \to h + v t$. Let me focus again on $d=1$ for a while. 
There is a more fundamental way to arrive to Eq. \eqref{KPZ}, which is to
start from the EW equation and write all possible terms which are local, allowed by symmetry, and relevant. Note that since one describes growth, the $h\to-h$ symmetry is broken
and the non linear term in \eqref{KPZ} is allowed. From power counting at the EW fixed point,
$h \sim x^{1/2} \sim t^{1/4}$ one sees that this term is relevant at large scale since $(\partial_x h)^2 \sim x^{-1} \gg \partial^2_x h \sim x^{-3/2}$. By the same counting argument higher order non-linearities
are irrelevant, e.g. 
$(\partial_x h)^4 \sim x^{-2} \ll \partial^2_x h$, so we can 
omit these terms when expanding $\sqrt{1+ (\partial_x h)^2}$. 
The equation \eqref{KPZ} will thus generically describe stochastic growth.
Since the non-linearity is relevant it will lead to a new universality class
with different growth exponents. It is intrinsically non-equilibrium since, contrarily to the EW equation, it cannot be written as a derivative of some Hamiltonian, i.e. there is no energy functional. So it will describe an irreversible process such as the growth of a stable phase into an unstable one.


The first property of the KPZ equation which was emphasized in \eqref{KPZ}
is its relation to the Burgers equation (which is the Navier-Stokes equation without pressure for an irrotational flow).
This relation is valid in any $d$, and written here in $d=1$ for
simplicity. Defining a velocity field 
$v = - \partial_x h$ and taking a derivative of \eqref{KPZ} one obtains
\be 
\partial_t v + \lambda v \partial_x v = \nu \partial_x^2 v - \partial_x \eta \label{Burgers} 
\ee 
which is the stirred Burger equation. This equation 
appeared a few years before in a paper by Forster, Nelson and Stephen \cite{Forster},
about dynamical RG for stirred $d$ dimensional Navier-Stokes. 
It was (briefly) pointed out there that for Burgers in $d=1$ 
there are exact exponents,
because there are two symmetries. This was developed further in the KPZ paper. What are these two symmetries ? The first one is Galilean invariance, natural from fluid dynamics. 
For the interface, defining 
\be 
\tilde h(x,t)= h(x + \lambda \epsilon t,t) + \epsilon x + \frac{\lambda \epsilon^2}{2} t 
\ee 
it states that if $h(x,t)$ is a solution of \eqref{KPZ} then $\tilde h(x,t)$ is also a solution
(with a tilted white noise, but which has the exact same statistics than the original noise). 
This symmetry is also called the statistical tilt symmetry (STS). Fixing $x,t$ one can
choose $\epsilon=-x/(\lambda t)$ and one obtains an equality in probability
\be 
h(x,t) \underset{\text{\rm in law}}{=}  h(0,t) - \frac{x^2}{2 \lambda t} 
\ee
Hence $h$ scales as $x^2/t$ which implies a first exponent relation, $\chi=2-z$.
This relation holds in any space dimension $d$. What is special about $d=1$ is that there
is a second exponent relation. It arises because the stationary probability distribution of the
KPZ equation in $d=1$ can be obtained exactly, and coincides with the one of the EW equation
${\cal P}_{\rm stat}[h] \sim e^{- \frac{\nu}{2 D} \int dx (\partial_x h)^2}$. This is a priori far from
obvious since the KPZ equation is non-equilibrium. A (very) heuristic argument is that the divergence of the 
probability current associated with the non-linear term 
$\sim \int dx \frac{\delta}{\delta h(x)}  (\partial_x h)^2 {\cal P}_{\rm stat}[h]  
\sim  \int dx (\partial_x h)^2 \partial_x^2 h$
is a boundary term in $d=1$ (e.g. it vanishes on a ring) \cite{ParisiBrownian}. It means that if
you generate an initial height field at time $t=0$ as a Brownian motion (BM) in space, 
it will again be distributed as a BM in space at any time $t$ \footnote{strictly, only the height differences $h(x,t)-h(0,t)$ become stationary, the "zero mode" is linearly growing and retains non trivial statistics,
see e.g. \eqref{onepoint} below}. 
This was noted
for stirred Burgers in \cite{Forster}, alluded to in \cite{KPZ}, with more details in \cite{hhf_85}.
Mathematicians have
proved that property \cite{Bertini,Funaki,Hairer} (and 
recently provided a rigorous version \cite{QuastelGuStationary}
of the argument in \cite{ParisiBrownian}).  
This implies that $\chi=1/2$ which in turns implies $z=3/2$, and the two scaling 
exponents are thus known in $d=1$. Interestingly, similar conclusions were reached in the related context of a lattice gas where the
density field obeys $\partial_t \rho + \partial_x j=0$, the current $j$ being a non-linear 
function of $\rho$, and accounting for additional diffusion and noise \cite{Spohn1985}.
For $j \sim \rho^2$ this also leads to the Burgers equation for $\rho$, hence
to the KPZ equation upon space integration.

The other beautiful property of the KPZ equation emphasized in \cite{KPZ} 
 is its connection to directed polymers in random media, an equilibrium stat mech problem involving quenched disorder (for review see e.g. \cite{hh_zhang_95}). This is done through the Cole-Hopf transform.
Starting from \eqref{KPZ}
and defining $Z(x,t)=\exp( \frac{\lambda}{2 \nu} h(x,t))$, the new field $Z(x,t)$
satisfies a linear equation, with multiplicative noise (here again in $d$ dimension)
\be \label{SHE}
\partial_t Z = \frac{T}{2} \nabla^2 Z - \frac{V(x,t)}{T} Z 
\ee 
where we set $T=2 \nu$ and $V(x,t)=-\lambda \eta(x,t)$. It is
often called the stochastic heat equation (SHE) in the math literature,
and looks like a Schrodinger equation in imaginary time.
Using the Feynman-Kac formula, its solution with initial condition 
$Z(x,0)=\delta(x-y)$, denoted $Z(x,t|y,0)$ can be written as a path
integral
\be 
Z(x,t|y,0) = \int_{x(0)=y}^{x(t)=x} {\cal D} x(\tau) 
e^{- \frac{1}{T} \int_0^t d\tau [ (\frac{dx(\tau)}{d\tau})^2 + V(x(\tau),\tau) ] } 
\ee 
which can be interpreted as the canonical partition sum at temperature $T$ of a continuum directed polymer in dimension $d+1$,
with fixed endpoints at $(y,0)$ and $(x,t)$. The first term in the exponential can be seen 
as an elastic energy, and the second as the potential energy collected along the polymer path
in a quenched random potential $V(x,\tau)$. The first term thus penalizes deviations from a vertical path, 
while the second encourages the polymer path to visit favorable regions where $V<0$.
The competition leads to a rough polymer, and the prediction from the KPZ scaling exponents is thus that for long polymers (corresponding to large time) the
polymer wandering scales as $x(\tau) \sim \tau^{1/z}$, which is 
$x(\tau) \sim \tau^{2/3}$ in $d=1$, i.e. superdiffusion \cite{numer2}. Although this
is a finite temperature model, long polymers are localized close to the optimal path, which corresponds to some zero-temperature fixed point, see below.
Conversely, the KPZ height field in a given noise
configuration 
is $h(x,t)= T \log Z(x,t)$, i.e. minus the free energy of the polymer in a given sample of the
random potential. So the solution to the KPZ equation in dimension $d$ can be represented by a directed polymer problem in dimension $d+1$, a powerful mapping. 
Alternatively the free energy of a directed polymer can be thought as resulting from a growth process
when its (projected) length $t$ increases.

Let me now briefly discuss the KPZ equation \eqref{KPZ} 
in higher dimension. As found in the KPZ paper \cite{KPZ}, $d=2$ is
a critical dimension. Indeed, at the Gaussian
fixed point (EW equation), $h \sim x^{\frac{2-d}{2}}$ hence $\nabla^2 h \sim x^{-1-\frac{d}{2}}$ and $(\nabla h)^2 \sim x^{-d}$, and the non-linearity in \eqref{KPZ}  is relevant at large scale only for $d<2$. Defining the dimensionless strength of the non-linear term, $g=\frac{S_d}{(2 \pi)^d} \frac{\lambda^2 D}{4 \nu^3} \Lambda_\ell^{d-2}$, they derived its RG flow (near $d=2$) 
\be 
\partial_\ell g = \beta(g) = (2-d) g + g^2 + O(g^3) 
\ee 
where $\Lambda_\ell = \Lambda e^{-\ell}$ is the running UV cutoff.
Hence for $d \leq 2$ the nonlinear coupling $g$ flows to large values
into the strong noise phase, which cannot be accessed perturbatively.
But for $d > 2$, 
there is an intermediate fixed point, at $g=g^* \simeq d-2$,
which separates a weak noise phase where the nonlinearity is irrelevant, and the strong noise phase. The bifurcation at $d=2$, described by this RG, is only about the appearance of a distinct weak noise phase (i.e. a high temperature phase for the directed polymer) for $d>2$. It does not tell much about the putative strong coupling KPZ fixed point, which appears to exist in any dimension. The mapping to the directed polymer gives some
information about this fixed point: it lies at $g=+\infty$, i.e. at zero temperature for the directed polymer. Numerical determination of the directed polymer ground state (optimal path) using a polynomial transfer matrix algorithm allows to obtain good estimates of the roughness/wandering exponent, $x \sim t^{\zeta}$, and of the sample to sample fluctuation of the optimal energy $E_0 - \overline{E_0} \sim t^\theta$ ($t$ denote the length of the polymer). 
Thanks to the Cole-Hopf mapping one has $\zeta=1/z$ and $\theta=\chi/z$. 
It is believed (and proved in $d=1$) that at this fixed point temperature is irrelevant, with $T_{\rm eff} \sim t^{- \theta}$, and the polymer path is localized around the ground state, up to rare large scale excitations (at  least within a "droplet" picture \cite{FisherHuse}). Thinking of the polymer as an elastic line, this strong disorder phase has the characteristics of a glass phase, with multiple local energy extrema and diverging energy barriers \cite{LarkinReview}. 

Concerning the KPZ scaling exponents in $d \geq 2$ there is little known beyond numerical simulations. Galilean invariance (i.e. the STS symmetry) ensures the relation $\chi=2-z$.
The conjecture $\chi=2/(d+3)$ was put forward by Kim and Kosterlitz, based on numerics on the restricted solid on solid (RSOS) growth model \cite{KimKosterlitz}. More recent numerics show that it is not exact, with, in $d=2$, $\chi = 0.3869(4)$ for
stationary RSOS interfaces \cite{PagnaniParisiNumerics}, and $\chi = 0.387-0.390$ for a variety
of models including directed polymers \cite{HalpinHealyNumerics} (see  \cite{HalpinHealyExperiments} for an experiment). Note also that there are some predictions, including universal amplitudes,
using non-perturbative RG \cite{Canet}.

At present some analytic solutions for $d>1$ are available only in the large $d$ limit. 
The first example is the exact solution of the directed polymer on the Cayley tree by 
Derrida and Spohn \cite{DerridaSpohn} and the second is the replica variational
ansatz of Mezard and Parisi \cite{MezardParisi} to describes the Gibbs measure of the directed polymer, which becomes exact for $d \to +\infty$. Both methods give insight in the large $d$ limit and arrive at the same qualitative conclusion, i.e. that in that limit there is a glass phase (the low temperature/strong disorder phase) with one step replica symmetry breaking phenomenology,
corresponding to $\theta=0$ (leading also to some insight into Burgers turbulence
\cite{BMPTurbulence1995}).
What is not known at present, and a matter of debate, is whether there is a finite upper critical dimension $d_{uc}<+\infty$ 
for the directed polymer and the KPZ equation, beyond which one would recover this mean-field behavior. 
In Ref. \cite{CookDerrida}
$1/d$ expansions were performed.
Another approach extends the polymer into an elastic manifold 
of higher internal dimension $D>1$, and uses the functional RG with $d$ components \cite{Wiese2loopN},
see \cite{PLDMullerCusps} for references and the connection 
with the replica method of \cite{MezardParisi} at large $d$.

Finally in $d=2$ one can ask whether there are signatures of 
conformal invariance in the statistics of e.g. stationary KPZ interfaces.
Recently it was shown numerically within a discrete KPZ dynamics on lozenge tilings \cite{Cao}
that KPZ level lines do not exhibit conformal invariance, contrarily to an earlier claim \cite{Saberi}, but that one could define another loop ensemble showing the
conformal invariance features of critical percolation. 

\section{A second KPZ revolution}

Let us now describe briefly the second KPZ revolution, around 2000,
with the solution of discrete models in the 1d KPZ class, called determinantal 
as their transition probabilities can be expressed as determinants, for which
the height distribution can be computed, unveiling connections to random matrix theory (RMT) and to random partitions. Before getting there allow me for a short but necessary digression. 

\subsection{RMT, determinantal processes and Tracy Widom distributions}

Let me briefly recall some basics about RMT, a priori unrelated topic (for more see \cite{Mehta,Forrester_book}).
Consider a $N \times N$ random matrix $H$ with centered Gaussian i.i.d entries
with the constraint of being either real symmetric ($\beta=1$) or hermitian ($\beta=2$),
and probability density $\propto e^{-\beta \frac{N}{4} {\rm Tr} H^2}$. Here $\beta$ is called the Dyson index. This density
being invariant under either rotations, or unitary transformation, the ensembles
are called Gaussian Orthogonal Ensemble (GOE) and Gaussian Unitary Ensemble (GUE) respectively. The third classical ensemble, for $\beta=4$, is the Gaussian Symplectic Ensemble (GSE). Their $N$ eigenvalues are real, and denoted 
$\lambda_i$. They are correlated random variables, and their joint PDF takes the form
$P[\lambda] \propto |\Delta(\lambda)|^\beta 
e^{-\beta \frac{N}{4}  \sum_{i=1}^N \lambda_i^2}$, where $\Delta(\lambda)=\prod_{i < j} (\lambda_i-\lambda_j)$ is the Vandermonde determinant. For any $\beta$, in the large $N$ limit the spectral density $\rho(\lambda)=\frac{1}{N} \sum_i \delta(\lambda-\lambda_i)$ converges to the Wigner semi-circle, with support $[-2,2]$.

Here we will be interested in the distribution of the largest eigenvalue $\lambda_{\rm max}=\max_i \lambda_i$ at large $N$. The order of its fluctuations can be guessed from the square root behavior of the spectral density near the upper edge, by 
considering the first quantile, $\frac{1}{N} \sim \int_2^{\lambda_{\rm max}} (2-\lambda)^{1/2}$ which gives $2 - \lambda_{\rm max} = O(N^{-2/3})$.
We will rescale $H \to N H$ so that the largest eigenvalue behaves at large $N$ as
\be \label{lambdamax} 
\lambda_{\rm max} \simeq 2 N + \chi N^{1/3}  \quad , \quad {\rm Prob}(\chi<s) = F_\beta(s) 
\ee 
where $\chi$ is a random variable. Its distribution for $\beta=1,2,4$ was obtained by Tracy and Widom (TW) \cite{TW-GUE,TW-GOEGSE}. Its cumulative distribution function (CDF) is denoted $F_\beta(s)$,
and the corresponding random variable is often denoted $\chi_\beta$. It is quite different from the Gumbel distribution, 
as the eigenvalues are highly correlated. Amazingly, anticipating a bit, the
exponent $1/3$ in \eqref{lambdamax} will turn out to be related to the KPZ growth exponent. 

To describe the TW distributions let me focus on $\beta=2$. In that case that the joint PDF
$P[\lambda]$ can be written as a determinant
$P[\lambda] \sim \det_{1 \leq i,j \leq N}  K(\lambda_i, \lambda_j)$ where $K(\lambda,\lambda')$ is called the kernel. Such a random point process (i..e the set of $\lambda_i$'s) is called determinantal and obeys that the marginal PDF of any number $n<N$ of eigenvalues 
is itself a (smaller) determinant involving the same kernel, $\sim \det_{1 \leq i,j \leq n}  K(\lambda_i, \lambda_j)$. This is akin to Wick's theorem for fermions, and indeed
it is easy to see that for $\beta=2$, $P[\lambda]$
coincides with the Slater determinant which describes the ground state of $N$ non-interacting fermions (at positions $x_i \propto \lambda_i$) 
in a harmonic trap. The kernel $K$ is then the projection operator on the space spanned by the $N$ lowest energy eigenfunctions of the harmonic oscillator. Now the important property
of determinantal processes is that the probability that there is no particle in some subset $J$ of the real line, called the hole probability, is equal to a Fredholm determinant defined as
\be \label{FDexpansion} 
{\rm Det}(1- K_J) := \sum_{n \geq 0} \frac{1}{n!} \int_{J^n} \prod_{i=1}^n d\lambda_i 
\det_{1 \leq i,j \leq n}  K(\lambda_i, \lambda_j)
\ee 
equivalently defined as usual from the eigenvalues of the linear operator $K_J$, where $K_J$ is the restriction of the kernel $K$ to $J$.
Choosing $J=[\lambda_{\max},+\infty[$
one can thus express the CDF of the largest eigenvalue as ${\rm Prob}( \lambda_{\max} < \lambda) = {\rm Det}[I - K_{[\lambda_{\max},+\infty[
}]$. In the large $N$ limit, $K$ converges, up to some rescaling called the edge scaling,
to the Airy kernel $K_{\rm Ai}$, so that finally the GUE-TW distribution is given by
the Fredholm determinant
\be 
F_2(s)= {\rm Det}[I - P_s K_{\rm Ai} P_s ] \quad , \quad 
K_{\rm Ai}(a,b)=\int_0^{+\infty} dv {\rm Ai}(a+v) {\rm Ai}(b+v) 
\ee 
where $(P_s K_{\rm Ai} P_s)(a,b)=\theta(s<a) K_{\rm Ai}(a,b)  \theta(s<b)$.
It is very easy to compute with a few lines of Mathematica, using Bornemann's
approximations by finite size determinants with exponential convergence \cite{Bornemann}.
There is an alternative formula, 
$F_2(s)=\exp( - \int_s^{+\infty} dt (t-s) q(t)^2)$ where $q(t)$ is solution of the Painleve equation $q''=t q + 2 q^3$ with $q(t \to +\infty) \sim {\rm Ai}(t)$. 
For $s \to -\infty$ it has a cubic lower tail $F_2(s) \sim e^{-|s|^3/12}$, and 
a stretched exponential lower tail $F_2(s) \sim e^{- \frac{4}{3} s^{3/2}}$ for $s \to +\infty$.
It turns out that
the GOE-TW distribution has also a simple determinantal formula \cite{FerrariSpohnGOE},
$F_1(s)={\rm Det}[I- B_s]$, with the kernel $B_s(a,b)=\theta(a) {\rm Ai}(a+b+s) \theta(b)$,
and the alternative formula \cite{TW-GOEGSE} 
$F_1(s)= F_2(s)^{1/2} \, e^{- \frac{1}{2} \int_s^{+\infty} q(t) dt}$.

\begin{figure}[h]
     \includegraphics[width=0.8\columnwidth]{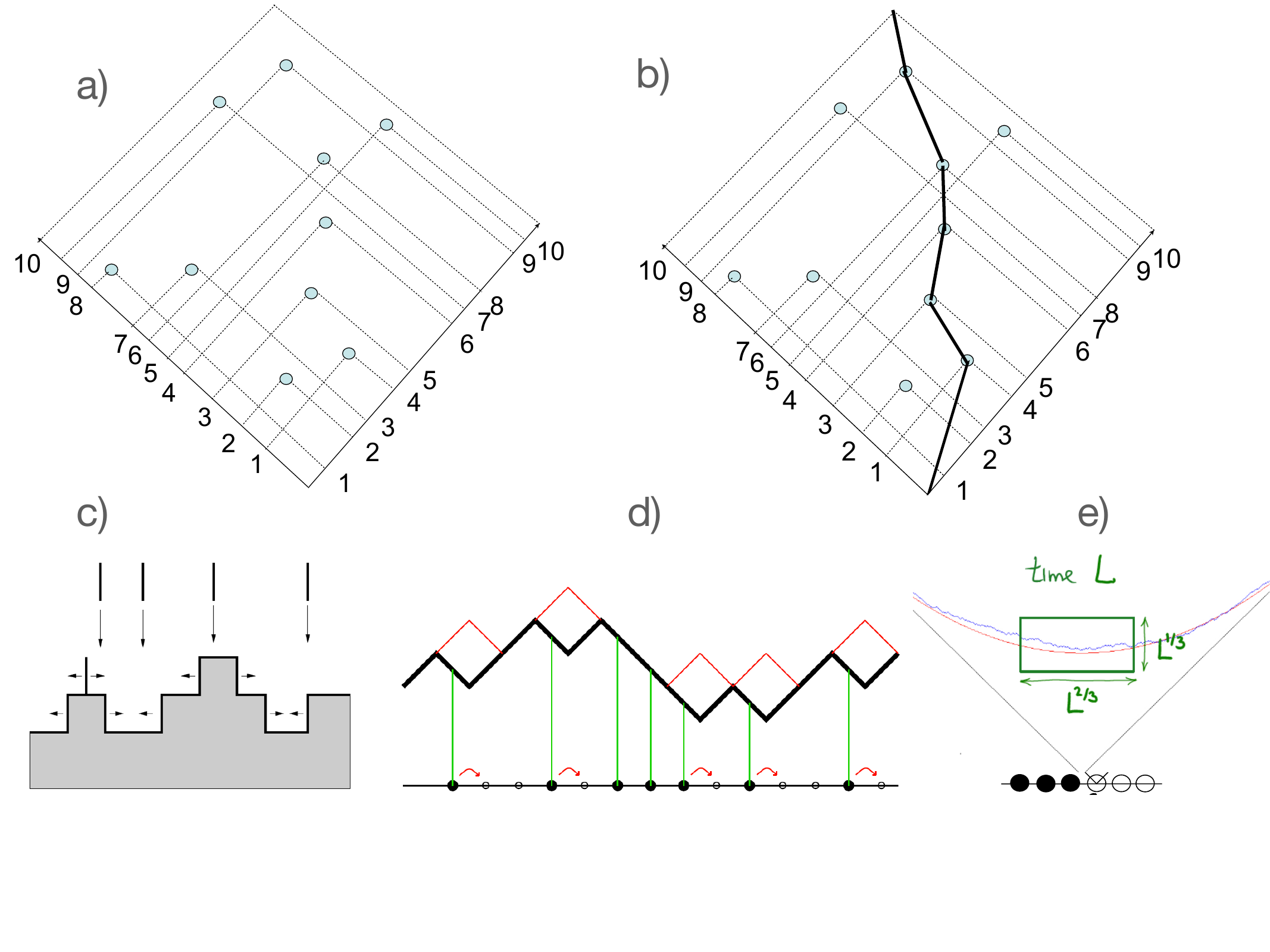}
   \vspace{-1.6cm}
   \caption{a) $N=10$ Poisson points in a square and the associated random permutation
   $\{8,2,7,\underline{1},\underline{3},\underline{4},10,\underline{6},\underline{9},5 \}$
   b) The LIS (underlined) ${1,3,4,6,9}$ corresponds to the optimal path going up along each axis and
   collecting the most points c) Polynuclear growth model d) totally assymetric exclusion process 
   e) evolution of TASEP from a "wedge" initial condition (also called
   "droplet")}
 \label{fig:LIS}
\end{figure}

\subsection{One point distributions} Let us return to the second
KPZ revolution, which started with the solution in combinatorial mathematics 
of Ulam's problem: what is the length of the
longest increasing subsequence (LIS) in a random permutation? Take the
permutation $\{8,2,7,1,3,4,10,6,9,5 \}$ with $N=10$ elements. One
sees that one LIS is $1,3,4,6,9$, of length $\ell_N=5$ (there is another one of the same
length). Baik, Deift and Johansson 
\cite{BaikDeiftJohansson}
showed that for large size of the permutation $N \gg 1$ it grows like $\sqrt{N}$ plus GUE-Tracy Widom fluctuations, i.e.  $\ell_N \simeq 2 \sqrt{N} + \chi_2 N^{1/6}$. 
This is based on a quite amazing formula, which gives the probability 
of this length in terms of a matrix integral on the unitary group,
${\rm Prob}(\ell \leq n) = e^{-\lambda} \int_{U(n)} dM e^{\sqrt{\lambda} {\rm Tr} (M + M^+)}$.
This matrix integral was studied previously in high energy physics \cite{GrossWitten,PeriwalShevitz}.
Note that $\ell$ in that formula is over a different ensemble, where $N$ is distributed
with a Poisson distribution of parameter $\lambda=\langle N \rangle$.

Why is this relevant for KPZ growth? Amazingly, Ulam's problem can be mapped
onto finding the optimal energy for a directed polymer in a random environment.
Indeed, let us throw points randomly in a square as a Poisson process,
as shown in Fig. \ref{fig:LIS} a). For a given realization, one orders the $x$ coordinates of these points from $1$ to $N$ ($N=10$ in the figure). Ordering their $y$ coordinates as well then defines a random permutation from the $x$'s labels to the $y$'s labels. A LIS then maps exactly to an "optimal" path going up (along both axis) and collecting the maximum possible number of points, as shown in Fig. \ref{fig:LIS} b). 
This brings the GUE-TW
distribution to the realm of directed polymers, for a special zero temperature discrete model of "point to point" 
directed polymer. Related results were
obtained by Prahofer and Spohn for the polynuclear growth model (PNG)
\cite{PrahoferSpohnPRL2000,Prahofer-Spohn2002}. In that model particles fall randomly on a surface:  each particle then nucleates a terrace, which grows laterally at unit speed, and terraces merge when they meet,
see Fig. \ref{fig:LIS} c). This process mimics the growth of an Ising interface in presence of an external field. 
It was thus shown that the fluctuations of the scaled height of the PNG process 
follow a Tracy-Widom distribution at large time.
Another closely related model is the totally asymmetric exclusion process (TASEP), a paradigmatic model in 
interacting particle transport in $d=1$. Particles on $\mathbb{Z}$ move at random times one unit to the right but only if the destination site is not occupied. To each particle configuration one associates a height field (increasing by one unit at a site if no particle is present, decreasing if there is
a particle), see Fig. \ref{fig:LIS} d). The dynamics of the TASEP is thus equivalent to the growth of a surface (particles moving to the right leads to a growing height). Johansson studied the case where
all particles are initially on the left of $0$, which for the height corresponds to a "wedge" initial condition, also called "droplet" or "curved" because the profile
becomes curved at later time. He showed \cite{JohanssonShape2000} that 
the (properly rescaled) height profile at large time has fluctuations again described by the GUE-Tracy-Widom distribution. This is illustrated in Fig. \ref{fig:LIS} e).
 Similar results for other models of directed polymers on the square lattice (also callled last passage percolation) also became available. Hence, a number of solvable models in the 1d KPZ class were found at that time (these models being "determinantal" and 
"at zero temperature"). 

In all these models one can define a "time" and a height field $h(x,t)$, and one finds that
at one point, say $x=0$, and at large time, the height grows as
\be \label{onepoint} 
h(0,t) \simeq_{t \to +\infty}  v_\infty t + (\Gamma t)^{1/3} \chi 
\ee 
where $v_\infty$ and $\Gamma$ are non-universal constants, and $\chi$ is a random variable. 
In the 1d KPZ class the distribution of $\chi$ belongs to a set of 
universal distributions, depending only on some broad features of the initial condition,
the initial finite scale features being erased by the growth.
For an initial height profile in the form of a wedge (or droplet/curved)  $\chi$ has a GUE-TW distribution,
while for a flat, or almost flat, initial profile it was shown that $\chi$ has
a GOE-TW distribution.

\subsection{Multi-point distributions} What about multi-space point height correlations? 
The 1d KPZ class is also characterized by universal multi-point distributions. 
Interestingly, it involves again random matrices, more precisely matrix Brownian motion. Let $H(\tau)$ be the (stationary) solution of the $N \times N$ matrix Ornstein-Uhlenbeck process
\be 
\partial_\tau H = - \frac{1}{N} H + W(\tau) 
\ee 
where $W(\tau)$ is a Hermitian matrix white noise ($\tau$ is some fictitious time). The largest eigenvalue $\lambda_1(\tau)$ is the top line of a $\beta=2$ stationary
Dyson Brownian motion (DBM) \cite{Mehta}.
At large $N$ it can be rescaled as a stochastic process in
the scaled variable $u=\tau/N^{2/3}$ 
\be 
\lambda_1(\tau) \simeq 2 N + N^{1/3} {\cal A}_2(\tau/N^{2/3}) 
\ee 
where $\{ {\cal A}_2(u) \}_{u \in \mathbb{R}}$ is called the Airy process, which is stationary and reversible. 
it is an extended determinantal process, meaning that there are explicit determinantal formula for
the joint PDF's of the form (e.g. for two points)
\be 
{\rm Prob}( {\cal A}_2(u_1) \leq  s_1, {\cal A}_2(u_2) \leq  s_2) = {\rm Det}_{2 \times 2}[\delta_{ij} I - P_{s_i} K_{u_i-u_j} P_{s_j})]
\ee 
where $K_u(a,b)=\pm \int_{v \in \mathbb{R}^\pm} dv e^{-v u} {\rm Ai}(a+v) 
{\rm Ai}(b+v)$ with $+$ for $\tau \geq 0$ and $-$ for $\tau<0$. 
At fixed $u$, the variable ${\cal A}_2(u)$, has a GUE-TW distribution. 
In \cite{Prahofer-Spohn2002,Johansson2003} it was shown that
for the PNG model as $x,t \to+\infty$ with fixed $\hat x$
\be 
(\Gamma t)^{-1/3} (h(x,t) - v_\infty t) \simeq {\cal A}_2(\hat x) - \hat x^2 
\quad , \quad \hat x=  \frac{A x}{2 (\Gamma t)^{2/3}} 
\ee 
where $A$ is the other non-universal constant.
This was obtained for the so-called PNG droplet, where nucleation occurs only in $[-t,t]$,
so that the interface develops a curved droplet shape ($h=0$ for $|x|>t$). 
In fact, this is true for all models in the KPZ class: for droplet/wedge/curved type initial conditions, the rescaled height profile converges at large time/distance to the
Airy$_2$ process minus the parabola. Equivalently, given the relation between the DBM and
Brownian motions, it says that the KPZ interface can be seen, in some window 
and under some rescaling, as the
top curve of a large set of non-crossing Brownian motions (also called watermelons). 

What about other class of initial conditions? For flat initial conditions, the scaled interface is
described by another process,
called Airy$_1$.  One difference with Airy$_2$ is that the two point height correlations decay as a power of 
the distance for
Airy$_2$, while they decay exponentially for Airy$_1$.

\subsection{Stationary KPZ} What about starting from a stationary interface ? (e.g. for TASEP
with i.i.d Bernoulli occupations). The (scaled) equal time height differences, as we already discussed,
should be distributed as a Brownian motion. But, the zero mode (e.g. the height at one
point) has a non-trivial distribution. 
In a pioneering work, Baik and Rains 
\cite{BR1} constructed the proper setting.
They considered again
2D Poisson points of rate unity in a square of side length $t$, but also with 1D Poisson points on the boundaries, with a different rate $\alpha$. For $\alpha>1$ it is
favorable for the optimal path (from $(0,0)$ to $(t,t)$) to stick for a finite fraction of its length 
to one of the
boundary before getting inside the bulk. The critical value $\alpha=1$ corresponds
to a stationary polymer \footnote{one can think of an infinitely long
polymer coming from $-\infty (1,1)$ and to its entry point inside the first quadrant}.
For $\alpha=1$ another interesting distribution emerges, and the total number of
points collected behaves as $\ell(t) \simeq 2 t + \chi_{\rm BR} t^{1/3}$ 
with ${\rm Prob}(\chi_{\rm BR} <s)=F_0(s)$ where $F_0(s)=\frac{d}{ds} [F_2(s)
\int_{-\infty}^s e^{- 2 \int_s^{+\infty} dt q(t)}]$. This is the so-called 
Baik-Rains (BR) distribution. 

In 2004 Prahofer and Spohn showed that for the
the PNG model with stationary initial condition, the scaled height fluctuations
obey the BR distribution. 
They also computed the space-time correlation of the height and of the Burgers velocity field $v \sim \partial_x h$ \cite{PrahoferSpohnStationary}
\be 
\langle (h(x,t)-h(0,0))^2 \rangle_c \simeq (\Gamma t)^{2/3} g( \frac{A x}{ 2 (\Gamma t)^{2/3}}) \quad , \quad \langle v(x,t) v(x',t') \rangle \propto |t-t'|^{-2/3} f_{\rm KPZ}( \frac{A |x-x'|}{2 (\Gamma |t-t'|)^{2/3}} ) 
\ee 
with $g(y) \simeq 2 |y|$ at large $y$, to be consistent with $h(x)-h(0)$ 
behaving as $\sqrt{A}$ times the standard Brownian on large scales. Here 
$f_{\rm KPZ}(y)=\frac{1}{4} g''(y)$ with 
$f_{\rm KPZ}(y) \sim e^{-c |y|^3}$ at large $|y|$. This universal function
for space-time stationary correlations is a hallmark of the 1d KPZ class, and
as we will see below, appears ubiquitously. 

For completeness, let us mention early exact solutions based on vertex models
and asymmetric spin Hamiltonians \cite{GwaSpohn,Kim1995,Dhar1987},
to which we return below.

\begin{figure}[t]
     \includegraphics[width=1.0\columnwidth]{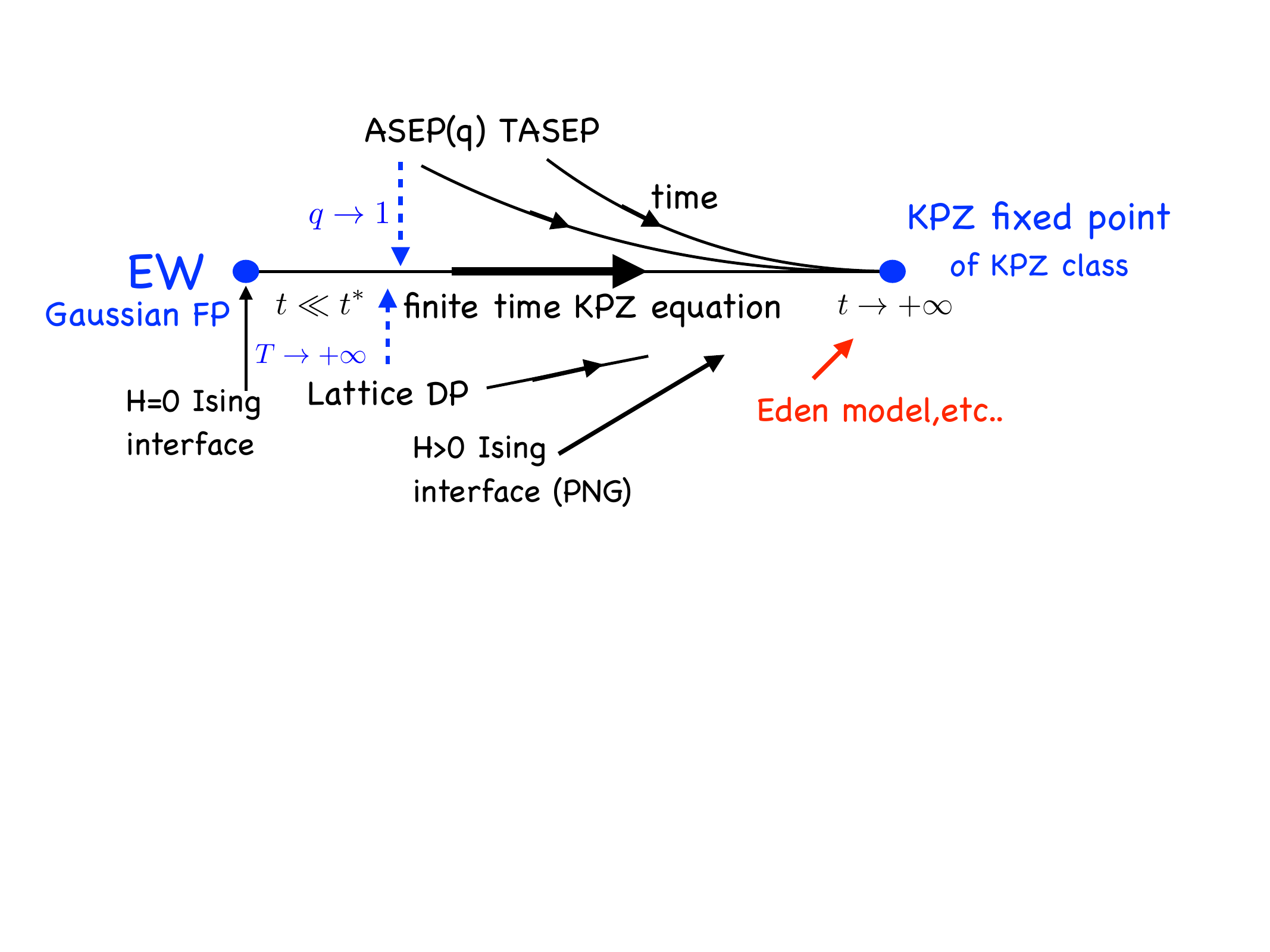}
     \vspace{-5.6cm}
   \caption{Flow of models in the 1d KPZ class (solvable or not) as time increases. The finite time KPZ equation is a flow line from EW to the KPZ fixed point. Many (e.g. finite temperature) models can be scaled to this flow line, for any fixed $t$, this is weak universality. All models in the KPZ class can be scaled at large time to the KPZ fixed point, this is strong universality. The solvable models allow a complete characterization of the fixed point. } 
 \label{fig:Flow}
\end{figure}

\subsection{KPZ equation, weak and strong universality}

Let us return to the continuum KPZ equation \eqref{KPZ}, and focus on $d=1$, space-time white noise, and its solutions for all times. Natural units are $x_0=(2 \nu)^3/(2 D \lambda^2)$, $t_0=x_0^2/\nu$ and $h_0=2 \nu/\lambda$. Everywhere below we will use these units, which amounts to fix $\nu=D=1$, $\lambda=2$.
For the KPZ equation the parameter $A$ is fixed by the stationary measure and equals
$A=D/\nu$, so it becomes $A=1$ in these units.
 \footnote{Let us stress that everywhere we use the convention of the seminal paper \cite{KPZ}, i.e. 
$\eta=\sqrt{2D} \, \xi$ in \eqref{KPZ}, where $\xi$ is the unit space
time white noise, while $\eta=\sqrt{D} \, \xi$ is also used in a number of papers}. For short time $t \ll 1$ the nonlinear term is not important, and
one recovers the EW equation (as easiest seen with a flat initial condition). As $t$ increases one is supposed to reach the KPZ fixed point, but how? Is the KPZ equation itself in the KPZ class? And what does the KPZ equation at finite time describe?
To answer the last question, consider a variant of TASEP called the ASEP 
where now the particles can jump to the right with rate $p=1$, or the left with rate $q=e^{-\epsilon}$. 
 It was proved \cite{Bertini} 
 that for weak asymmetry $\epsilon \ll 1$, the field
 $Z_\epsilon(x,t)=\exp( \frac{\epsilon}{2} h(4 \epsilon^{-2} x, 16 \epsilon^{-4} t) +c t)$ 
 converges as $\epsilon \to 0$ to a solution of the continuum SHE. Here $x,t$ are fixed and there is no
 notion of large time. This property is called weak universality. Similarly if one considers the 
 recursion relation for a directed polymer on a lattice, and one takes the high temperature limit, under appropriate rescaling it also converges to the SHE/KPZ equation at finite times. 
 The situation is summarized in Fig. \ref{fig:Flow}. The finite time KPZ equation is thus an interpolation between the Edward-Wilkinson fixed point and the strong KPZ fixed point. The KPZ fixed Point attracts all the models in the KPZ class, and that is called strong universality. But for any fixed $t$ some models 
have the KPZ equation as a limit, and this is weak universality. Note that the "zero-temperature" models
cannot be scaled to the KPZ equation. 
 
\section{The 1d KPZ class in experiments} 
\begin{figure}[h]
     \includegraphics[width=0.47\columnwidth]{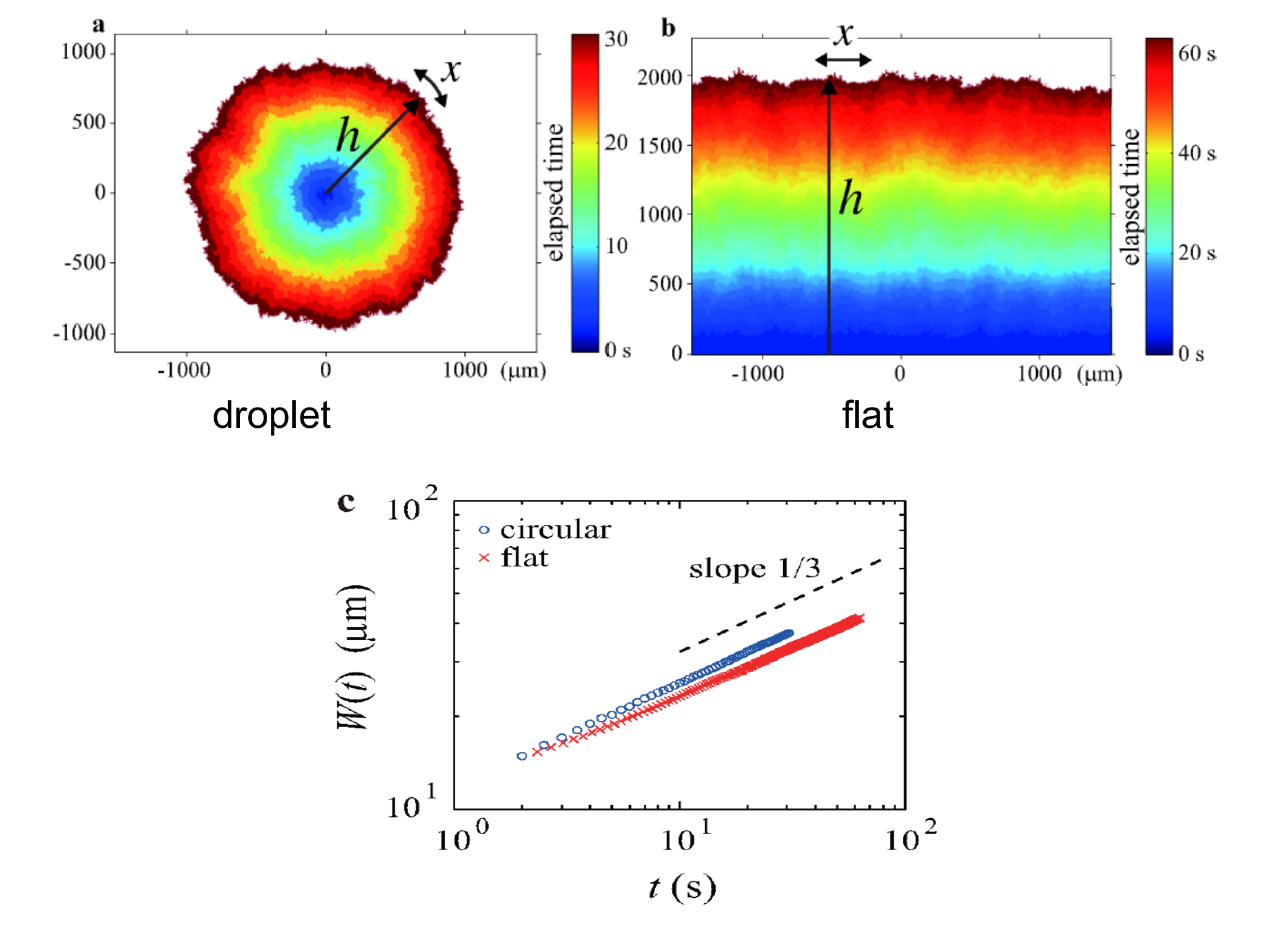}
      \includegraphics[width=0.6\columnwidth]{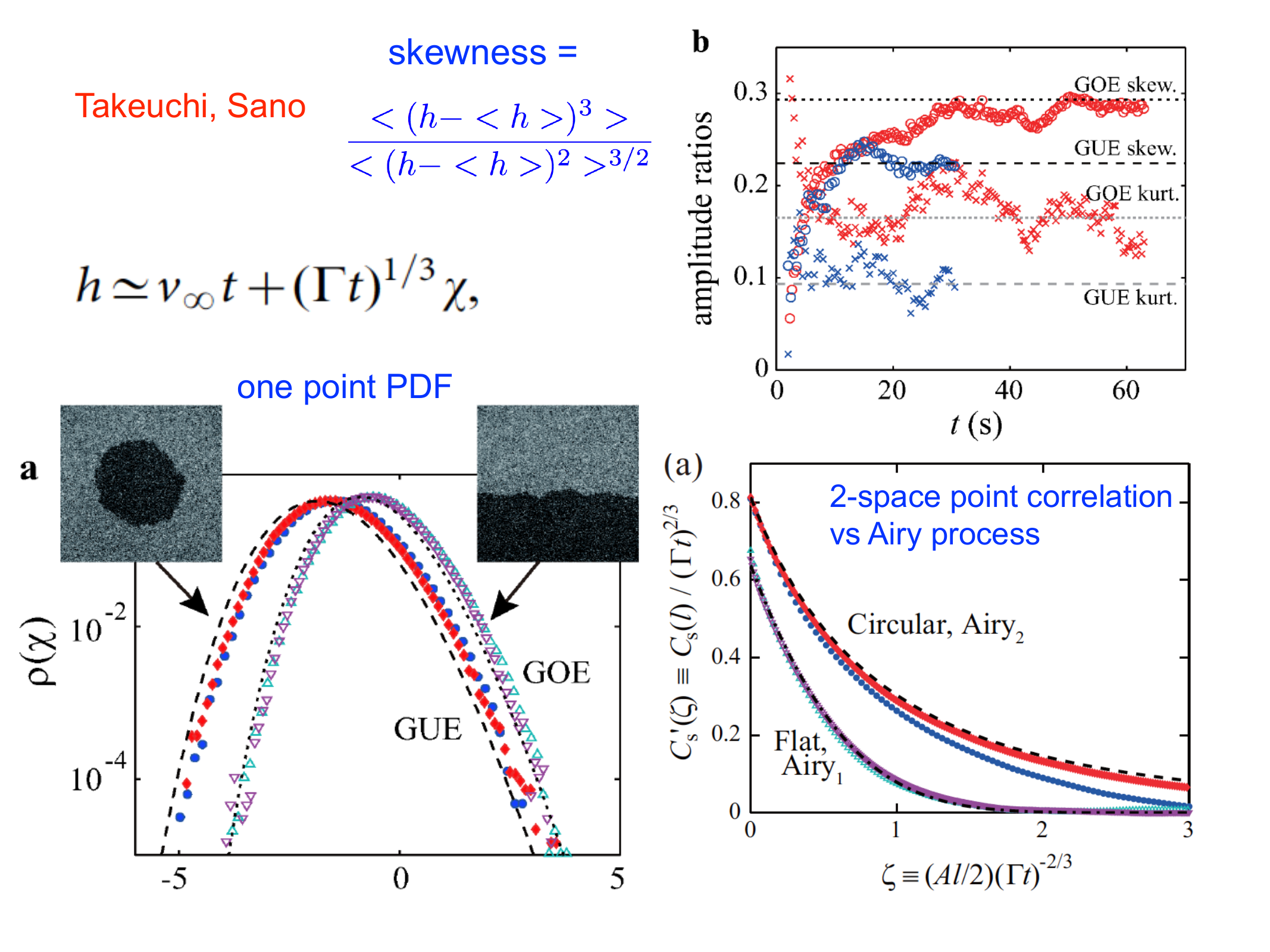}
   \caption{Liquid crystal experiments from \cite{Takeuchi1} and \cite{TakeuchiGeometry}. Left:
   interface growth from a point (droplet/curved) and from a line (flat), and interface width as a function of time.
   Right:  bottom left: histogram of the one-point height distribution of the scaled variable $\chi$, as compared to GUE and GOE Tracy-Widom distributions, top right:  convergence of skewness and kurtosis of $\chi$ towards their respective TW values, bottom right:  two point spatial correlation as
   compared to the predictions from the Airy processes, from \cite{TakeuchiGeometry}.}
 \label{fig:exp}
\end{figure}

Have these intricate predictions for the 1d KPZ class been tested in experiments ?
Surprisingly, even for the scaling exponents it took quite some time to obtain very clean evidence. 
The KPZ exponents were reported in the slow combustion of paper \cite{SlowCombustion},
in the growth of cell colonies \cite{CellColonies2010} in the formation of coffee rings via evaporation
\cite{CoffeeStains}, and more. A breakthrough was achieved by Takeuchi and Sano around 2010 with liquid crystals in a two-dimensional geometry
\cite{Takeuchi1,TakeuchiGeometry}. By shining a laser, they nucleated a stable phase inside an unstable phase. In Fig. \ref{fig:exp} the color is encoding time and one sees how the interface is growing,
either from a point into a circular droplet (the so-called curved or droplet initial condition) 
or from a line (flat initial condition). Remarkably, they can create any initial condition, recently they even created a Brownian initial interface to test stationary KPZ
\cite{TakeuchiBR}. They observed the 1d KPZ growth exponent $\beta=1/3$,
and many more very clean signatures of the 1d KPZ universality class. But since KPZ
is so ubiquitous, why did it take so long? The reason is that it is not so easy to avoid quenched disorder or long-range effects in experiments. Takeuchi and Sano used turbulent liquid crystals, so that all effects of disorder are washed out, and each experimental run is very fast, which allows to collect huge statistics.
Thanks to this improved statistics they were able to show very good agreement of the PDF of the scaled variable $\chi$ in \eqref{onepoint} with the GUE and GOE TW distributions, down to the tails, as well as agreement for the skewness and kurtosis. As also shown in Fig. \ref{fig:exp}, the comparison of the spatial two point height 
correlation with those of the Airy processes are excellent (with consistent values of the non-universal parameters $A$ and $\Gamma$). These experiments stimulated the community, and experimental tests of the 1d KPZ class have been obtained in other systems, including in condensed matter physics, see Section \ref{sec:applications} below.

 
\section{A third KPZ revolution, or KPZ Renaissance} 

Around 2010 some new developments occured in physics, and many more in mathematics, about the KPZ equation and the KPZ class in $d=1$. 
 
\subsection{Replica solutions of the KPZ equation}

It turns out that it is possible to "solve" the continuum KPZ equation using replicas, and the method is called the "replica Bethe ansatz". One first maps it to the continuum directed polymer in a random potential, and one uses replica on that stat mech problem with quenched disorder. The replicated polymer problem can be recast in terms of quantum bosons in $d=1$ in imaginary time. For white noise, one obtains the so-called delta Bose gas, which can be solved using the Bethe ansatz. This route was pioneered by Kardar around 1987
\cite{Kardar87} (see also \cite{ParisiBrownian,KardarNelsonIncommensurate1985}). A pending question has been how to extract from it the PDF of the KPZ height. That second part was only achieved around 2010
\cite{LeDoussal1,Dotsenko1,Dotsenko2}, let me briefly explain how.

{\bf Droplet initial condition}. Let me focus on a polymer with fixed endpoints, corresponding to the so-called narrow wedge/droplet initial condition for the KPZ equation. We start by computing integer moments of the partition sum, $\overline{Z^n} = \overline{e^{n h}}$, 
and from them we will extract the distribution of the height $h$.
This is illegal because one finds that the moments grow much too fast with $n$ to uniquely determine the distribution, but we
put aside mathematical rigor for now.
The key property is that the averages of products of $n$ partition sums with different endpoints 
${\cal Z}_n := \overline{Z(x_1,t) \dots Z(x_n,t)}$ satisfy an evolution equation 
(an exercise in Ito calculus starting from the SHE)
\be
\partial_t {\cal Z}_n = - H_n  {\cal Z}_n \quad , \quad H_n = - \sum_{i=1}^n \partial_{x_i}^2 
- 2 \sum_{1 \leq i \leq j \leq n} \delta(x_i-x_j) \label{LL} 
\ee 
which involves the Hamiltonian of $n$ quantum particles with an attractive delta interaction (attraction between replica always arise from quenched disorder). 
The $n$-th moment of the partition sum 
of the point to point directed polymer, $Z=Z(x_0 t| x_0 0)$,
can
thus be written as 
\be 
\overline{Z^n} = \langle x_0 \dots x_0 |e^{-t H_n}| x_0 \dots x_0 \rangle
= \sum_\mu |\Psi_\mu(x_0,\dots,x_0)|^2 e^{- t E_\mu} \label{sum}
\ee 
where we introduced the eigenfunctions $\Psi_\mu$ and eigenenergies $E_\mu$
of $H_n$, and
the symmetry of the observable implies that only bosonic states contribute.
For the Lieb-Liniger model \eqref{LL} the eigenstates are known from the Bethe ansatz \cite{Lieb-Liniger,McGuire,Yang,gaudin,HeckmanOpdam1997} and
are linear superpositions of plane waves with phase factors. They are indexed by a set of $n$ complex rapidities, 
which are solutions of the Bethe equation. Since we have $n$ bosons with attraction it leads to bound states. The ground state has a very simple form 
$\Psi_0(\vec x) \sim \exp( - \frac{1}{2} \sum_{i<j} |x_i-x_j|)$ and 
energy $E_0=- \frac{1}{12} n (n^2-1)$. Assuming that it dominates
at large time led to the prediction that
$\overline{Z^n}  \sim e^{\frac{1}{12} n^3 t}$,
a very fast growth with $n$ \cite{Kardar87}. Although 
the combination $n^3 t$ is suggestive of a 
$t^{1/3}$ free energy scaling exponent,
one cannot really continue this formula in $n$, since the double limit of infinite space and time has been taken, which does not commute with the limit $n \to 0$ \cite{Brunet-Derrida}.
Instead, these integer moments give insight into the large deviation
upper tail of the distribution of the height field \cite{LargeDevPLD2016}, ${\cal P}(h) \sim_{ h \sim t \gg 1} e^{- \frac{4}{3} (h/t)^{3/2} t}$. One can check that it recovers 
$\overline{e^{n h}} = \int dh e^{n h} {\cal P}(h) \sim e^{\frac{1}{12} n^3 t}$ 
from a saddle point at $h/t = n^2/4$. It matches the
upper tail of the GUE-TW distribution 
$F'_2(s) \sim_{s \to +\infty} e^{- \frac{4}{3} s^{3/2}}$ which describe the (much smaller) typical height
fluctuations, $h/t^{1/3}=s=O(1)$, but, surprisingly, it retains the same form.

Hence, to obtain the PDF of the height $h$ at fixed time $t$, and then take $t$ large, one needs to first sum over all the eigenstates in \eqref{sum}.
Although formidable looking, it is in fact possible, as the excited states have 
a simple structure: they are obtained by partitioning the $n$ particles into $1 \leq n_s \leq n$ 
bound states called "strings", each with $m_j \geq 1$ particles and total momentum $m_j k_j$, with 
$n = \sum_{j=1}^{n_s} m_j$. The total eigenenergy of each such partition is simply the sum of the kinetic energy and binding energy of each string, 
$E_\mu=\sum_j m_j k_j^2 - \frac{1}{12} m_j(m_j^2-1)$. Inserting the eigenfunctions, 
from \eqref{sum} one then obtains an explicit formula for $\overline{Z^n}$ as a sum over partitions of multiple integrals over the $k_j$. Interestingly the next step is {\it not}
to try to continue analytically in $n$ this complicated expression and take $n \to 0$, but instead one introduces
a generating function (analogous to the one introduced in \cite{DerridaSpohn} for the
polymer on the Cayley-tree) 
\be \label{generating} 
g_t(s) = 
1 + \sum_{n \geq 1} \frac{(-e^{- t^{1/3} s})^n}{n!} \overline{\tilde Z^n}
= \overline{ \exp( - e^{- s t^{1/3} } \tilde Z ) } =
\overline{ \exp( - e^{- t^{1/3} (s - \tilde h)} ) } 
\ee 
It is constructed so that in the large time limit $g_\infty(s)=\lim_{t \to +\infty} g_t(s)
= {\rm Prob}(\tilde h<s)$ equals the CDF of $\tilde h$. Here $\tilde h$ is the properly shifted and scaled height 
\footnote{here $\tilde Z= Z e^{t/12}$ so that at large time $h=\log Z \simeq - \frac{1}{12} t + t^{1/3} \tilde h$ where $\tilde h=O(1)$}. 
The sum over $n$ in \eqref{generating} can now be reorganized as an expansion in
the number of strings, 
$g_t(s) = 1 + \sum_{n_s \geq 1} Z(n_s,s)$, where each factor $Z(n_s,s)$ 
is a free sum over the $m_j \geq 1$, $j=1,\dots,n_s$. There are several crucial ingredients. First one can tame the hopelessly diverging series $\sum_{m_j \geq 1} e^{\frac{1}{12} m_j^3 t}$ using an "Airy trick" based on the 
identity $\int_{-\infty}^{+\infty} dy Ai(y) e^{y w} = e^{w^3/3}$, converting it into a nicer geometric
series. Next, the factor $|\Psi_\mu(x_0,\dots,x_0)|^2$ for any string state turns out to be a Cauchy determinant. Hence a determinantal structure emerges and 
the sum over $n_s$ takes the form of a Fredholm determinant expansion \eqref{FDexpansion}, eventually leading to \cite{LeDoussal1,Dotsenko1,Dotsenko2,Kormos}
\begin{equation} \label{FDfinitetime} 
g_t(s) = {\rm Det}[ I + P_s \hat K_{t} P_s] \quad , \quad 
K_t(a,b) = \int_{-\infty}^{+\infty} du \frac{{\rm Ai}(a+u) 
{\rm Ai}(b+u)}{ 1 + e^{- u t^{1/3}} }
\end{equation}
Thus the generating function \eqref{generating} can be written for any time $t$ as
a Fredholm determinant in terms of a "finite time" kernel $K_t$. 
Clearly as $t \to +\infty$ this kernel converges to the Airy kernel $K_\infty=K_{\rm Ai}$,
which shows convergence of the CDF of $\tilde h$ to $F_2(s)$ the GUE-TW 
distribution (hence at the physics level of rigor the KPZ equation belongs to the KPZ class!). 
More precisely, this shows that \eqref{onepoint} holds for droplet initial conditions, with $\chi=\chi_2$ (and in these
units $\Gamma=1$). 

This finite time solution has interesting (not strongly-universal) features
specific to the KPZ equation itself.
First by Laplace inversion from \eqref{generating}, \eqref{FDfinitetime} one can
obtain an explicit expression (in terms of Fredholm determinants)
for the finite time PDF of $\tilde h$
\cite{LeDoussal1,KPZ-TW1c,ProlhacSpohnPDF2011}.
Second, the apparent "Fermi factor" in the
expression of $K_t$ (which arises from the geometric summation over the $m_j$) 
allows to map this problem \cite{DeanFiniteT} to
the quantum statistics of the rightmost fermion at the edge of a Fermi gas in a trap at a finite temperature
$T_{\rm fermions} \sim t^{-1/3}$. Finally, this finite time formula has interesting applications to study the large deviations of the PDF of the height.
At short time $t \ll 1$, one shows that the latter takes the form ${\cal P}(h) \sim \exp( - \frac{\Phi(h)}{\sqrt{t}} )$,
with a non-trivial rate function, $\Phi(h) \sim_{|h| \ll 1} h^2$ (consistent with $h \sim t^{1/4}$ typical EW fluctuations) and $\Phi(h) \sim_{h \to +\infty} |h|^{3/2}$,
$\Phi(h) \sim_{h \to -\infty} |h|^{5/2}$ for very atypical fluctuations
\cite{LargeDevPLD2016,Meerson2016,Kraj}.
At large time $t \gg 1$, the lower tail takes the form
${\cal P}(h) \sim \exp( - t^2 \Phi_-(h/t))$ (we discussed the upper tail above)
where the non-trivial rate function $\Phi_-$ can be obtained using Coulomb gas methods
\cite{SasorovLargeTime,CorwinPLD,Kraj},
akin to those developped in RMT, thanks to the (surprising) determinantal structure
of the finite time solution.
It matches the TW tail $\Phi_-(z) \sim z^3$ for $z \to 0^-$ but behaves as $|z|^{5/2}$ for $z \to -\infty$.
\\

{\bf Other initial conditions}. There are only a few initial conditions for which the one-point distribution of the height of the KPZ equation has been obtained analytically at arbitrary time
\footnote{\eqref{sum} can be generalized, but requires the analytical knowledge of the overlaps
of the initial condition ${\cal Z}_n(t=0)$ with the Bethe states}. 
For the flat initial condition $Z(x,0)=1$, $g_t(s)$ is obtained using replicas as a Fredholm Pfaffian, i.e. its square is
the Fredholm determinant of a 2 by 2 matrix kernel
\cite{PLD-Flat1,PLD-Flat2}. For Brownian initial conditions, 
i.e. $Z(x,t=0)=e^{B(x)}$, a Fredholm determinant formula was also
obtained using replicas for a variant of $g_t(s)$ \cite{ImamuraStat}, with an Airy type kernel, deformed by Gamma functions (a cousin of the Baik-BenArous-Peche (BBP) transition kernel \cite{BBP} for outliers of spiked
random matrices). In both cases, it showed convergence to the GOE-TW and the BR distribution at large time.
Replica solutions were also obtained for the KPZ equation on a half-line, i.e. a directed polymer on a half space, in which case the third classical ensemble of RMT, the GSE, appears.
\\

\subsection{Rigorous solutions and developments in mathematics}

This coincided with an explosion of works in probability theory and mathematical physics,
a field often called integrable probability, or stochastic integrability, at the interface between statistical mechanics 
and representation theory/integrable systems. Clearly the replica Bethe ansatz solutions, working directly on the continuum KPZ equation, are non-rigorous. Another route, which allows for rigorous results, is
to identify integrable discrete stochastic models in the KPZ class, and
to perform at the end the continuum limit towards the KPZ equation. Of course
these discrete models are also interesting in their own right. 
One model which was found early on
to be integrable is the ASEP, for which Tracy and Widom obtained Bethe ansatz type contour integral formula \cite{TWASEP1}. Using these formula and performing the weak asymmetry limit (using the aforementioned weak universality) led to rigorous derivations \cite{KPZ-TW1a,KPZ-TW1b,KPZ-TW1c,AmirCorwinQuastel} (in parallel and at about the same time), of the
finite time solution \eqref{FDfinitetime} of the KPZ equation with droplet initial condition.

It is important to note that the ASEP is "solvable" in a different sense as the TASEP.
The ASEP has an integrable structure, and allows a limit to the finite time KPZ equation.
The TASEP is determinantal (the transition probabilities can be expressed as
determinants - the so-called Schutz formula \cite{Schutz1997}) 
and has in common with the KPZ equation
only its large time universal behaviour (strong universality). 
A similar distinction holds for polymer models.
While the last passage percolation models are zero-temperature determinantal models, and often directly
relate to random matrix ensembles, new models were discovered, such as the log-gamma polymer and its semi-discrete version, the O'Connell-Yor polymer, which are finite temperature polymer models, integrable in the Bethe ansatz sense. 

{\bf A tree of models}. At that time an impressive pyramid/tree
of integrable stochastic models was discovered \cite{MacDo}
(see also \cite{CorwinReview} for a review and Fig. 9 there), as a 
probabilistic application of representation theory and symmetric function theory.
At the top of the tree
lie the so-called Macdonald processes, which are probability measures/Markov processes 
on interlacing triangular arrays 
$\lambda_j^{(m)}$, $1 \leq j \leq m$,
(Gelfand-Tsetlin patterns), which can be seen as interacting particle systems. 
They depend on two parameters $q,t$ and they degenerate in various successive limits of these parameters to various other integrable systems. For instance one leaf of the tree of models contains the Gaussian $\beta$ random matrix ensembles (the rows $\lambda_j^{(N)}$ scaling to the $N$ eigenvalues). Another one, for $t=0$, contains the so called $q$-Whittaker processes, for which the side of the triangle $\lambda_j^{(j)}$ maps to
the positions of particles in the so called $q$-TASEP. The $q$-TASEP is a $q$-deformation of the TASEP (obtained for $q=0$) 
such that particles jump only to the right, with rate $1-q^{\text{gap}}$, 
where $q<1$ and $\text{gap}$ is the number of empty sites before their right nearest neighbor. 
Taking further $q \to 1$ leads to the aforementioned integrable finite temperature polymer models
and eventually to the KPZ equation itself. For all these discrete models one can derive nested 
contour integrals "moment formula" \cite{MacDo}, e.g. in the $q$-TASEP 
for observables schematically of the type 
$\overline{q^{n h}}$ where $h$ is some definition of a "height field". One sees that
now the PDF of $h$ is uniquely defined from its moments, since $0<q<1$,
which allows rigorous results for the KPZ equation in the limit $q \to 1$.
This is the so-called "rigorous replica" approach. It led, among many other things, to a rigorous
derivation of the finite time stationary solution \cite{Veto} obtained from replica in \cite{ImamuraStat}.
Interestingly though, the replica solution for the flat initial condition is still not completely proved, although
there is little doubt that it is correct \cite{QuastelFlatASEP2015}. 

The pyramid of models got even richer around 2015 with the exploration of
stochastic vertex models in mathematics. Vertex models, and the Yang-Baxter equations, 
played an important role in the development of integrable models
in a wide variety of fields in physics and mathematics
\cite{BaxterBook,FaddeevHouches,Reshetikhin}. In some cases,
they can represent stochastic dynamics \cite{GwaSpohn}
and analogs to moment formula can be obtained \cite{BCG6Vertex2014}.
The family got even richer with introduction of higher spin vertex models \cite{CorwinPetrovHigherSpinVertex2016} 
and colored vertex models \cite{BorodinWheeler2018} associated to 
higher rank quantum affine groups. In the replica Bethe ansatz framework
the latter models are associated to the nested Bethe ansatz, and
allow to adress questions related to the "Airy sheet", see below.

{\bf The KPZ fixed point}. The Cole-Hopf formula gives the solution of the
KPZ equation at time $t$ in terms of the initial data, as $e^{h(x,t)} = \int dy Z(x,t|y,0) e^{h(y,t=0)}$, where $Z(x,t|y,0)$ the partition sum of the directed
polymer starting at $(y,0)$ and ending at $(x,t)$. In the universal large time limit the variations of $h$ 
grows as $t^{1/3}$ and the integral is dominated by the optimal path and optimal initial position $y$.
The statistics of the height field at one point $x$ can then be set a variational
problem involving the Airy process
\footnote{To obtain the scaled height as a process in $x$, 
one should replace ${\cal A}_2(\hat x-\hat y) \to {\cal A}_2(\hat x, \hat y)$,
the Airy sheet, see below} 
\be \label{variational} 
t^{-1/3} h(x,t) \simeq \max_{\hat y} [ {\cal A}_2(\hat x-\hat y) - \hat y^2 + h_0(\hat y) ] 
\ee
and the rescaled initial condition (IC) $h_0(\hat y)=t^{-1/3} h(2 t^{2/3} \hat y,0)$. 
For flat IC $h_0(\hat y)=0$, for droplet $h_0(\hat y)=-\infty$ except for $\hat y=0$ where $h_0(0)=0$,
and for Brownian IC $h_0(\hat y)=\sqrt{2} B(\hat y)$. Solving such variational
problem for general initial condition is difficult but was achieved \cite{QuastelKPZFixedPoint}.
The $n$ point joint PDF of  ${\sf h}(x,t)$, a suitably rescaled height field
with universal statistics, was obtained as a $n \times n$ Fredholm determinant involving a kernel which
can be constructed for a general initial data in terms of hitting probabilities
of a Brownian
motion on the initial height profile. This universal statistics (any model in the KPZ class can
be rescaled to it) is called (in mathematics) "the KPZ fixed point", and was obtained from TASEP determinantal formula.
Amazingly, it was also shown \cite{QuastelKP} that at the KPZ fixed
point the the height PDF for general initial data, satisfies the Kadomtsev-Petviashvili equation
(an integrable 2D generalisation of the KdV equation, which describes non linear wave motion)
\be
F(t,x,r) = {\rm Prob}({\sf h}(x,t)<r) \quad , \quad \phi=  \partial_r^2 \log F 
\quad , \quad \partial_t \phi + \frac{1}{2} \partial_r \phi^2
+ \frac{1}{12} \partial_r^3 \phi + \frac{1}{4} \partial_r^{-1} \partial_x^2 \phi = 0
\ee 
Interestingly, no finite time multi-point replica solution is available for the KPZ equation,
however, under an (uncontrolled) "decoupling assumption" it is possible to obtain large time
multi-point formula (see \cite{QuastelKPZFPReplica} and references therein).

\subsection{Multi-time correlations and memory, Airy sheet and directed landscape}
What about the joint distribution of the KPZ height at different times ?
E.g. for for two times $t_1,t_2$, of $h(0,t_1)$ and $h(0,t_2)$.
When both times are large, under proper rescaling it depends only on the ratio $u=t_2/t_1>1$
and should be universal. Within the replica Bethe ansatz, starting from the KPZ equation,
it is notably difficult to calculate \cite{Dotsenko2time}, but a partial solution (when
the height at the earliest time, $h(0,t_1)$, is large) 
was obtained \cite{deNardisPLD2time} for droplet
initial conditions, and tested in experiments \cite{deNardisPLDTakeuchi2time}.
An interesting quantity is the universal dimensionless ratio 
$R= \lim_{u \to +\infty} \lim_{t_1 \to +\infty} 
{\rm Cov}(h(0,t_1),h(0,u t_1))/{\rm Var}(h(0,t_1))$.
Indeed if it is non zero it means that even at infinite time
the interface keeps a persistent memory of its initial growth (a kind of ergodicity breaking).
This happens in the droplet/curved geometry, for which this ratio was obtained analytically 
\cite{PLDAiry}, with $R \approx 0.623$ in agreement with numerics and experiments
\cite{TakeuchiGeometry,deNardisPLDTakeuchi2time,HH}. Notably, for stationary
profiles, it was possible to obtain the universal two time scaled height covariance \cite{FS2timeStat}
based only on general considerations about Airy processes. 

In mathematics the complete universal two time joint PDF for droplet
initial conditions was obtained recently 
(using strong universality) from lattice polymer models \cite{Johansson2times}, 
and from the TASEP on a ring \cite{BaikMultiTime}. The resulting formula are
quite heavy, and involve contour integrals of complicated Fredholm determinants,
and much work remains to analyze their properties. 

Finally, the statistics of the so-called Airy sheet ${\cal A}_2(\hat x, \hat y)$ remains
a challenge, despite recent progress. It is defined as the (scaled) optimal energy of polymer paths as a joint function of their (scaled)
endpoints, a universal object. A fruitful and paralllel line of research opened recently,
aiming to characterize (in a more abstract and general sense) the probability
measure on the full set of optimal paths with arbitrary endpoints
\cite{ViragDirectedLandscape2022,Virag2020,KPZeqDirectedWu2023}. These optimal 
paths can be seen as geodesics for some sort of random (directed) metric,
and there is a thus a (random) geometric
component as well.


%

I

\subsubsection{A fields medal around the KPZ equation}

Besides these developments focused on the large scale
behavior of the KPZ equation, there was another remarkable line of 
research in mathematics on its short scale, ultraviolet structure. It was pioneered by M. Hairer, and led to his 2014 Field's medal. It allowed to give a rigorous mathematical meaning to the $d=1$ KPZ equation
as a stochastic partial differential equation (SPDE). Indeed,
in the white noise limit Eq. \eqref{KPZ} leads to $h(x,t)$ which is locally Brownian in $x$,
for which the non-linear term $(\partial_x h)^2$ becomes ill-defined.
In physics language, the associated MSR dynamical field theory
has ultraviolet divergences which require renormalisation
to lead to a proper continuum limit. What Hairer did is to construct a probabilistic renormalisation
procedure directly for the non linear SPDE's
\cite{HairerRough,HairerKPZ,HairerRegularity}. 
In a sense Hairer achieved for SPDE's what Ito did for ODE.
To quote the IMU \cite{IMU} 
"Building on the rough-path approach of Lyons for stochastic ordinary differential equations, 
Hairer created an
abstract theory of regularity structures for stochastic partial differential equations (SPDEs).
This allows Taylor-like expansions around any point in space and time. The new
theory allowed him to construct solutions
to singular non-linear SPDEs as fixed points of a renormalization procedure.
Hairer was thus able to give, for the first time, a rigorous intrinsic meaning to many SPDEs 
arising in physics". And to prove the existence of
solutions to the KPZ equation at all times \footnote{these
are compatible with the Cole-Hopf solutions of the KPZ equation, $h = \log Z$,
where $Z$ is solution to the 
SHE - which can be rigorously defined \cite{Bertini}}. 
(and to its extensions with more general
non-linearities \cite{HairerQuastel}).

\section{Selected applications} \label{sec:applications} 

The applications of the KPZ equation encompass four decades and many fields, and are too numerous
to even start to summarize here. Let us focus on some notable examples of recent interest, where
the KPZ physics arises in a sometimes surprising way.
\\

{\bf Universal distributions of conductance}. The Anderson model $H= \sum_i \epsilon_i c^\dagger_i c_i - 
t \sum_{\langle i j \rangle} (c^\dagger_i c_j  + h.c.)$ 
with random site energies $\epsilon_i$ exhibits eigenfunction localization in space dimension $d=2$.
The KPZ class arises as follows. Within the locator expansion, the 
Green's function at fixed energy $G(x,y,E)$ can be written as a sum over paths from $x$ to $y$ of
products of weights $t/(E-\epsilon_i)$ along the path. In the deeply localized regime 
$|x-y| \gg \xi$, where $\xi$ is the localisation length, it was argued that directed paths dominate the
sum, and it becomes a $d=1+1$ directed polymer problem with, however, non positive (and in a magnetic field, complex) weights. The latter problem was further argued to belong to the 1d KPZ class \cite{MedinaKardar}. As a result it was 
predicted and checked numerically \cite{Ortuno}
that the conductance between two leads separated
by $L \gg \xi$ behave as 
\be 
\log g = - \frac{2 L}{\xi} + \alpha (\frac{L}{\xi})^{1/3} \chi
\ee 
where $\chi$ is a random variable, consistent with the GUE-TW distribution for narrow leads.
Further non trivial predictions from the 1d KPZ class were tested \cite{Lemarie}, as well as for 
semi-infinite systems \cite{SomozaPLD}. An even more surprising connection to
the 1d KPZ class was obtained recently in the related context of localization in 2D random Dirac operators
\cite{ViragGraphene,BarraquandVirag} 
(such as those relevant e.g. for electrons in disordered graphene sheets).

{\bf Driven interfaces and quenched KPZ equation}. 
An interface driven by an external force $F>0$ over a substrate with quenched disorder exhibits a non-linear velocity-force characteristics $v(F)$, with a depinning transition $v(F_c)=0$. The equation of motion of the interface 
takes the form \eqref{KPZ}, where the noise term is replaced by a quenched random force $\eta(x,t) \to \eta(x,h(x,t)) + F$. The KPZ term $\lambda$ in \eqref{KPZ}, even if absent in the microscopic model, is generated by motion
\cite{KesslerLevine1991,KardarLines,MovingGlass}.
This is the famous quenched KPZ equation \cite{ParisiQKPZ} (qKPZ). At large $v$ the
quenched noise becomes equivalent to white noise again and the moving interface (its roughness and dynamics) is described by the standard (thermal) KPZ equation. Much less obvious
is what happens for $v \to 0$, i.e. near depinning. It turns out that there are two
universality classes \cite{Anisotropic}, quenched EW and qKPZ, depending on whether the
KPZ term is absent or present for $v=0^+$, with different critical exponents (in turns, it depends
on whether the system is isotropic or not). Strikingly, in $d=1+1$
the qKPZ class is related to directed percolation \cite{DirectedPerco}. Importantly, for the qKPZ equation the
{\it sign} of the KPZ term matters, with facetting occuring in the negative $\lambda$ case: this can be oberved in experiments on chemical reaction fronts in porous media, which
nicely show three different universality classes \cite{Atis}.

{\bf Coupled KPZ equations: driven lines and crystals, non-linear fluctuating hydrodynamics, and 
the puzzling anomalous
spin transport in Heisenberg magnets}. An elastic object such as a vortex line or a dislocation
line in 3D, or a vortex lattice in 2D or 3D, is described by a multi-component displacement
field. When driven in a medium with quenched disorder, it naturally leads to a description
in terms of coupled, multicomponent, and often spatially anisotropic, KPZ equations \cite{ErtasKardar1992,KardarLines,Hwa,Ramaswamy1997,MovingGlass}. 
The RG analysis of these coupled KPZ equations (usually around $d=2$) is complicated, as found by Ertas and Kardar for the $n=2$ component system \cite{ErtasKardar1992}, and in most cases flows to strong coupling phases
(as the usual KPZ equation) for which little is known a priori
\footnote{one well studied exception is the anisotropic (single component) KPZ equation in $d=2$ 
with opposite sign non linearity $(\partial_x h)^2 - (\partial_y h)^2$, which flows
to the 2D Gaussian free field with logarithmic roughness for the height.}. Multicomponent KPZ (and Burgers) equations reappeared more recently in the context of 
non-linear fluctuating hydrodynamics (NLFH) in one space dimension, in the form \cite{CoupledKPZSpohn2013}
\be
\partial_t h_\alpha = - v_\alpha \partial_x h_\alpha + \sum_{\beta,\gamma=1}^n G^\alpha_{\beta \gamma} 
\partial_x h_\beta \partial_x h_\gamma + \sum_{\beta=1}^n D_{\alpha \beta} \partial_x^2 h_\beta 
+ \sum_{\beta=1}^n B_{\alpha \beta} \xi_\beta 
\ee 
where $\phi_\alpha= \partial_x h_\alpha$ are $n$ conserved fields. For instance it 
describes the coarse grained dynamics of 1D anharmonic chains for $n=3$ (number, momentum, and
energy are conserved) with two sound modes $v_{\pm 1}=\pm c$ and one
heat mode $v_0=0$ \cite{SpohnHydroReview}. In that case it was shown \cite{CoupledKPZSpohn2013} that in the steady state the phonon peaks take the scaling form
$S_\alpha(x,t)=\langle \phi_\alpha(x,t) \phi_\alpha(0,0) \rangle \simeq (\lambda_\alpha t)^{-2/3} 
f_{\rm KPZ}( (\lambda_\alpha t)^{-2/3}(x- v_\alpha t))$, $\alpha=\pm 1$, where $f_{\rm KPZ}$ is
the Prahofer-Spohn stationary KPZ scaling function described above. The standing heat mode may have a scaling function given by asymmetric L\'evy distributions \cite{SpohnStoltz2015}. 
A classification for any $n$, using mode coupling theory, was obtained  \cite{PopkovSchutz2015,PopkovModeCoupling2016,PopkovSchutzGolden2024}. Surprisingly, the Fibonacci numbers
$C_n=1,1,2,3,5,\dots$ appear, and the dynamical exponent is given by the discrete infinite set of
Kepler ratios $z_n = C_{n+1}/C_n$. More precisely, consider the mode $\alpha$:
(i) if $G^\alpha_{\beta \beta}=0$ for all $\beta$, then the mode $\alpha$
is diffusive, $z_\alpha=2$, and $S_\alpha$ is Gaussian (EW class),
(ii) if $G^\alpha_{\alpha \alpha} \neq 0$, then $z_\alpha=3/2$, and $S_\alpha$ is Prahofer-Spohn
or modified KPZ, and 
(iii) if $G^\alpha_{\alpha \alpha} = 0$, but $G^\alpha_{\beta \beta} \neq 0$ for some $\beta \neq \alpha$, then
$z_\alpha=3/2,5/3,8/5,13/8,\dots \omega$, and $S_\alpha$ is Levy.
The NLFH has an expanding range of applications, e.g. reaching to quantum field theories \cite{Delacretaz}. 


Recently, the KPZ physics made a very surprising apparition in the high temperature transport properties
of integrable 1D quantum (and classical) spin systems with non-Abelian symmetry.
Consider for instance the quantum Heisenberg spin $1/2$ chain
$H=\sum_i (S_i^x S_i^x + S_i^y S_i^y) + \Delta S_i^z S_i^z$. For 
$\Delta<1$ the spin transport (as probed by 
the spin-spin spacetime correlations) is ballistic, while it is diffusive for $\Delta>1$. 
At the isotropic point $\Delta=1$ it is found to be super-diffusive with characteristic
length $\xi(t) \sim t^{2/3}$ 
\cite{Prosen2017}. More strikingly, the high temperature 
equilibrium space-time spin-spin correlation was found numerically to be accurately described 
by the Prahofer-Spohn stationary KPZ scaling function 
\cite{Prosen2019}. While the $2/3$ exponent can be 
explained by scaling arguments involving large magnons 
\cite{VasseurScaling2019,deNardisBurgers}, it is much more mysterious
why the 1d KPZ class would appear here. It was soon
also observed in experiments, neutron scattering \cite{MooreNature2021}
and cold atoms \cite{Bloch2022}, and in numerics of various isotropic integrable 
spin chains, quantum or classical
\cite{Simulations}. Furthermore, this manifestation of the KPZ class is believed to be universal \cite{Universal},
the ingredients being integrability and isotropy. 
After some original attempt \cite{Bulchandani2020} 
to show the emergence of a 1D one-component noisy Burgers equation for the
magnetization $S^z \sim m \sim \partial_x h$, 
an improved two component theory taking into account the 
conserved energy current in the system was formulated in terms 
of two coupled noisy Burgers equations
\cite{JacopoCoupled2023} (similar, interestingly enough, to 
the aforementioned Ertas-Kardar equations for elastic lines \cite{ErtasKardar1992}). 
In that case
the two modes are standing $v_0=0$ (degenerate flux), and
the aforementioned classification does not apply \cite{SpohnDegenerate2024}.
In any case, this is likely not the ultimate theory as it does not seem to give the correct values for higher moments of the current fluctuations, computed in numerical simulations.
Therefore, 
at present, the situation is extremely puzzling. 
It seems that
two point spin correlations do belong to the one-component 
KPZ class, as recently tested numerically with great accuracy on the 1D Landau-Lifshitz
magnet 
\cite{Takeuchi2024}, but that higher order correlations do not (e.g. the kurtosis
is much smaller than predicted by the Baik-Rains distribution, or even half of it, 
as predicted by the two-mode theory). The field thus awaits a more complete
theory taking into account the particular nature of fluctuations in integrable models.

{\bf KPZ and quantum entanglement growth in random circuits}.
As was discovered recently \cite{Nahum2017} 
the KPZ physics appears in 
the dynamics of quantum information. 
Random circuits are much studied toy models 
for quantum chaotic systems. They are chains of quantum spins each with
a local Hilbert space of dimension $q$. At discrete times a
random unitary in $U(q^2)$ acts on a pair of adjacent spins
at a random location, which entangles them. Defining $\rho_x(t)$ the reduced density matrix at time $t$,
where all sites $x'<x$ are traced over, one define the 
Renyi entropies $S_n(x,t)=\frac{1}{1-n} {\rm Tr} \rho_x(t)^n$ which
measures the entanglement between the two half-systems.
Starting from a product state, the average entanglement entropy grows 
linearly with time. In the limit $q \gg 1$ it was shown \cite{Nahum2017}  that 
$S_n(x,t)$ evolves (for all $n$) as the height in a single step surface growth model
\cite{Meakin1986,KimKosterlitz} which is in the KPZ class. 
The conclusion that $S_n(x)$ obeys the KPZ equation (in a coarse grained sense)
holds for finite $q$ as well \cite{Nahum2019}, indicating that the fluctuations of the (overall linear)
entropy growth
are in the KPZ class, with e.g. $h(0,t)= S(0,t) \simeq v_E t + (\Gamma t)^{1/3} \chi$
at large time, where $\chi$ has a Tracy-Widom distribution. In follow up works \cite{Nahum2018} the
KPZ class was also found in operator spreading and out-of-time-order correlator
(OTOC) dynamics of chaotic systems. Interestingly, it also appears in some random integrable systems, 
but at the level of large deviations
\cite{BernardPLD2020}.

{\bf Superfluid condensate dynamics and polaritons}.
The KPZ physics is predicted to emerge also in the dynamics of superfluid condensates, but with a twist.
There the phase field plays the role of the KPZ height, and effectively obeys a KPZ equation
in any dimension
\cite{Grinstein1993,AltmanToner2015,DiehlSuperfluid2015}. The twist is
that the phase is periodic in $[0,2 \pi]$, and the compactness of the height field
leads to regimes where standard KPZ should hold, but also to different regimes, where topological defects proliferate
\cite{DiehlKPZVortex2017}. Nevertheless, the signatures of the 1d KPZ class
were recently observed experimentally in 1D condensate of polaritons in semiconductor cavities
\cite{PolaritonNature2022}, by carefully filtering for (mostly bound) vortex-antivortex pairs.  
The KPZ class also shows up in other systems with phase dynamics, such as in synchronization of oscillator lattices \cite{CuernoSynchronization2024}.

To close this list let us mention a few other diverse applications such as
the growth of perturbations 
in chaotic coupled logistic maps
\cite{TakeuchiChaos2021}, diffusion in time dependent 
random environments \cite{RWRECorwinBarraquand,TTPLD2017} 
and expansion range in bacterial colonies \cite{NelsonRangeExpansion2007}.
\\

{\bf Aknowledgments}. 
I am grateful to the organizers of the seminar series "The interdisciplinary contribution of Giorgio Parisi to theoretical physics", at the Sapienza University of Rome in 2022-2023, for inviting me to give a talk in the series, and for collecting in a volume the write-ups of all talks. I thank Maria Chiara Angelini for useful comments on this manuscript and Enzo Marinari for general guidance. I thank Giorgio Parisi for lively discussions, and sharing the
story of the birth of the KPZ equation. 

I also thank numerous colleagues for discussions along the years,
and apologize for the many references omitted by necessity in this short overview.


\begin{thebibliography}{12} 

\bibitem{KPZ} M.Kardar, G.Parisi, Y-C.Zhang, {\it Dynamic scaling of growing interfaces},
Phys.\ Rev.\ Lett.\ {\bf 56}, 889 (1986).


\bibitem{LarkinReview} For a review see,
Blatter, G., Feigel'man, M. V., Geshkenbein, V. B., Larkin, A. I., Vinokur, V. M. (1994). Vortices in high-temperature superconductors. Reviews of modern physics, 66(4), 1125.



    \bibitem{Witten} 
    Witten Jr, T. A., Sander, L. M. (1981). Diffusion-limited aggregation, a kinetic critical phenomenon. Physical review letters, 47(19), 1400
    
    \bibitem{QLE} 
    Jason Miller, Scott Sheffield, Quantum Loewner Evolution,
    arXiv:1312.5745, Duke Math. J. 165, no. 17 (2016), 3241-3378
    
\bibitem{KPZ0} 
V. G. Knizhnik, A. M. Polyakov, and A. B. Zamolodchikov. Fractal
structure of 2D-quantum gravity. Modern Phys. Lett. A, 3(8):819-
826, 1988. MR947880

\bibitem{KPZPolyakov}
A. M. Polyakov. From Quarks to Strings. ArXiv e-prints, November
2008, 0812.0183.

\bibitem{KPZProof} 
    B. Duplantier and S. Sheffield. Duality and the Knizhnik-PolyakovZamolodchikov relation in Liouville quantum gravity. Phys. Rev.
Lett., 102(15):150603, 4, 2009. 0901.0277. MR2501276 (2010d:83106).

\bibitem{footnoteKPZmeetKPZ} A recent workshop "KPZ meets KPZ" compared 
the statistics of geodesics in LQG
and in the so-called "directed landscape" associated to the KPZ class.
\url{http://www.fields.utoronto.ca/activities/23-24/kpzwork}.


\bibitem{EW} 
Edwards S F and Wilkinson D R 1982 Proc. R. Soc. A 381 17-31. 

\bibitem{SpohnHouches2016} 
H. Spohn, 
{\it The Kardar-Parisi-Zhang equation - a statistical physics perspective} 
arXiv:1601.00499,  to appear, Les Houches Summer School July 2015 session CIV "Stochastic processes and random matrices", Oxford University Press

\bibitem{FamilyVicsekEdenBallistic85}
Family, F., Vicsek, T. (1985). Scaling of the active zone in the Eden process on percolation networks and the ballistic deposition model. Journal of Physics A: Mathematical and General, 18(2), L75.

\bibitem{Eden2} 
Plischke, M., Racz, Z. (1985). Dynamic scaling and the surface structure of Eden clusters. Physical Review A, 32(6), 3825.

\bibitem{Forster} 
D. Forster, D.R. Nelson and M.J. Stephen 
Phys. Rev. A, 16 (1977) 732.


\bibitem{ParisiBrownian}
G. Parisi, On the replica approach to random directed polymers in two dimensions, 
Journal de Physique, 51 (1990) 1595.


\bibitem{hhf_85} D.A.\ Huse, C.L.\ Henley, and D.S.\ Fisher,
Phys.\ Rev.\ Lett.\ {\bf 55}, 2924 (1985).

\bibitem{Bertini}
Bertini L. and Giacomin G., 
Stochastic burgers and KPZ equations from particle systems, 
Comm. Math. Phys., 183 (1997) 571-607.

\bibitem{Funaki} 
Funaki T. and Quastel J., KPZ equation, its renormalization and invariant measures, 
Stoch. Partial Diff. Eq.: Anal. Computat., 3 (2015) 159.

\bibitem{Hairer} 
Hairer M. and Mattingly J., The strong Feller property for singular stochastic PDEs, , 54 (2018) 1314.

\bibitem{QuastelGuStationary}
Gu, Y., Quastel, J. (2024). Integration by parts and invariant measure for KPZ. arXiv preprint arXiv:2409.08465.

\bibitem{Spohn1985} 
van Beijeren, H., Kutner, R., H. Spohn (1985). Excess noise for driven diffusive systems. Physical review letters, 54(18), 2026.


\bibitem{hh_zhang_95} T.\ Halpin-Healy and Y-C.\ Zhang,
Phys.\ Rep.\ {\bf 254}, 215 (1995).

\bibitem{numer2} M.\ Kardar and Y-C.\ Zhang,
Phys.\ Rev.\ Lett.\  {\bf 58}, 2087 (1987).

\bibitem{FisherHuse}
Fisher, D. S., Huse, D. A. (1991). Directed paths in a random potential. Physical Review B, 43(13), 10728.


\bibitem{KimKosterlitz}
J.M. Kim, J. M., Kosterlitz,  (1989). Growth in a restricted solid-on-solid model. Physical review letters, 62(19), 2289.

\bibitem{PagnaniParisiNumerics}
Pagnani, A., G. Parisi, (2015). Numerical estimate of the Kardar-Parisi-Zhang universality class in (2+ 1) dimensions. Physical Review E, 92(1), 010101.

\bibitem{HalpinHealyNumerics}
T. Halpin-Healy, $2+1$-Dimensional Directed Polymer in a Random Medium: Scaling Phenomena and Universal Distributions. Physical review letters, 109(17), 170602 (2012),
and 
Extremal paths, the stochastic heat equation, and the three-dimensional Kardar-Parisi-Zhang universality class,
Physical Review E- Statistical, Nonlinear, and Soft Matter Physics, 88(4), 042118.

\bibitem{HalpinHealyExperiments} 
T. Halpin-Healy, G. Palasantzas, G. (2014). Universal correlators and distributions as experimental signatures of 2+ 1 Kardar-Parisi-Zhang growth. arXiv preprint arXiv:1403.7509.

\bibitem{Canet} 
Canet, L., Chat\'e, H., Delamotte, B., Wschebor, N. (2011). Nonperturbative renormalization group for the Kardar-Parisi-Zhang equation: General framework and first applications. Physical Review E-Statistical, Nonlinear, and Soft Matter Physics, 84(6), 061128.

\bibitem{DerridaSpohn} 
Derrida, B., Spohn, H. (1988). Polymers on disordered trees, spin glasses, and traveling waves. Journal of Statistical Physics, 51, 817-840.

\bibitem{MezardParisi}
M\'ezard, M., Parisi, G. (1991). Replica field theory for random manifolds. Journal de Physique I, 1(6), 809-836.


\bibitem{BMPTurbulence1995} 
Bouchaud, J. P., M\'ezard, M., Parisi, G. (1995). Scaling and intermittency in Burgers turbulence. Physical Review E, 52(4), 3656.

\bibitem{CookDerrida} 
Cook, J., Derrida, B. (1990). Directed polymers in a random medium: 1/d expansion and the n-tree approximation. Journal of Physics A: Mathematical and General, 23(9), 1523.

\bibitem{Wiese2loopN} 
Le Doussal, P., Wiese, K. J. (2005). Two-loop functional renormalization for elastic manifolds pinned by disorder in N dimensions. Physical Review E-Statistical, Nonlinear, and Soft Matter Physics, 72(3), 035101.


\bibitem{PLDMullerCusps} 
Le Doussal, P., Muller, M., Wiese, K. J. (2008). Cusps and shocks in the renormalized potential of glassy random manifolds: How functional renormalization group and replica symmetry breaking fit together. Physical Review B- Condensed Matter and Materials Physics, 77(6), 064203.


\bibitem{Cao} 
Cao, X., Rosso, A., Santachiara, R. (2015). Conformal invariance of loop ensembles under Kardar-Parisi-Zhang dynamics. Europhysics Letters, 111(1), 16001.



\bibitem{Saberi} 
Saberi A., Niry M., Fazeli S., Tabar M. R. and
Rouhani S., Phys. Rev. E, 77 (2008) 051607

\bibitem{Mehta} M.L.Mehta, 
{\it Random Matrices}, Elsevier, Amsterdam (2004)  



\bibitem{Forrester_book}
P. J. Forrester, {\it Log-gases and random matrices}, Princeton university press (2010).

\bibitem{TW-GUE} C.A.\ Tracy and H.\ Widom,
Commun.\ Math.\ Phys.\ {\bf 159}, 151 (1994)

\bibitem{TW-GOEGSE} 
C. Tracy and H. Widom, 
Comm. Math. Phys. 177:727-754 (1996).

\bibitem{Bornemann} 
Bornemann, F. (2010). 
On the numerical evaluation of Fredholm determinants. Mathematics of Computation, 79(270), 871-915.

\bibitem{FerrariSpohnGOE} 
P.L. Ferrari and H. Spohn, J. Phys. A 38 L557 (2005).


\bibitem{BaikDeiftJohansson} 
Baik, J., Deift, P., Johansson, K. (1999). On the distribution of the length of the longest increasing subsequence of random permutations. Journal of the American Mathematical Society, 12(4), 1119-1178.

\bibitem{GrossWitten}
Gross, D. J., Witten, E. (1980). Possible third-order phase transition in the large-N lattice gauge theory. Physical Review D, 21(2), 446.

\bibitem{PeriwalShevitz}
Periwal, V., Shevitz, D. (1990). Unitary-matrix models as exactly solvable string theories. Physical review letters, 64(12), 1326.


\bibitem{PrahoferSpohnPRL2000}
M. Prahofer, H. Spohn, 
Universal distributions for growth processes in 1+ 1 dimensions and random matrices, 
Phys. Rev. Lett. 84 4882 (2000).

\bibitem{JohanssonShape2000}
Johansson, K.: Shape fluctuations and random matrices. Commun. Math. Phys. 209, 437-476 (2000)

\bibitem{Prahofer-Spohn2002} M.\ Prahofer and H.\ Spohn
J.\ Stat.\ Phys.\ {\bf 108}, 1071 (2002)
Prahofer, M., Spohn, H. (2002). Scale invariance of the PNG droplet and the Airy process. Journal of statistical physics, 108, 1071-1106.

\bibitem{Johansson2003}
Johansson, K. (2003). Discrete polynuclear growth and determinantal processes. Communications in Mathematical Physics, 242, 277-329.

\bibitem{BR1} 
J. Baik and E.M. Rains, 
J. Stat. Phys.100 (2000), 523-542.

\bibitem{PrahoferSpohnStationary} 
Prahofer, M., Spohn, H. (2004). Exact scaling functions for one-dimensional stationary KPZ growth. Journal of statistical physics, 115(1), 255-279.

\bibitem{GwaSpohn}
L.-H. Gwa, H. Spohn, Six-vertex model, roughened surfaces, and an asymmetric spin Hamiltonian, Physical Review Letters, 68, no. 6 (1992), 725-728, and 
Bethe solution for the dynamical-scaling exponent of the noisy Burgers
equation, Phys. Rev. A 46, 844-854 (1992).

\bibitem{Kim1995} 
Kim, D. (1995). Bethe ansatz solution for crossover scaling functions of the asymmetric XXZ chain and the Kardar-Parisi-Zhang-type growth model. Physical Review E, 52(4), 3512.


\bibitem{Dhar1987} 
Dhar, Deepak. "An exactly solved model for interfacial growth." Phase transitions 9.1 (1987): 51.


\bibitem{SlowCombustion}
Maunuksela, J., Myllys, M., Kahkonen, O. P., Timonen, J., Provatas, N., Alava, M. J., Ala-Nissila, T. (1997). Kinetic roughening in slow combustion of paper. Physical review letters, 79(8), 1515.

\bibitem{CellColonies2010} 
Huergo, M. A. C., Pasquale, M. A., Bolzan, A. E., Arvia, A. J., Gonzalez, P. H. (2010). Morphology and dynamic scaling analysis of cell colonies with linear growth fronts. Physical Review E-Statistical, Nonlinear, and Soft Matter Physics, 82(3), 031903.


\bibitem{CoffeeStains}
 P. J. Yunker, M. A. Lohr, T. Still, A. Borodin, D. J. Durian,
and A. G. Yodh, Phys. Rev. Lett. 110, 035501 (2013)
Effects of Particle Shape on Growth Dynamics at Edges of Evaporating Drops of Colloidal Suspensions.
M. Nicoli, R. Cuerno, and M. Castro, Comment, Phys. Rev. Lett. 111,
209601 (2013).

\bibitem{Takeuchi1} 
K. A. Takeuchi and M. Sano, 
Phys. Rev. Lett. 104, 230601 (2010); 
K. A. Takeuchi, M. Sano, T. Sasamoto, and H. Spohn,
Sci. Rep. (Nature) 1, 34 (2011).


\bibitem{TakeuchiGeometry} 
K. A. Takeuchi, M. Sano, J. Stat. Phys. 147, 853-890 (2012).


\bibitem{TakeuchiBR} 
T. Iwatsuka, Y. T. Fukai, K. A. Takeuchi,
Phys. Rev. Lett. 124, 250602 (2020).



\bibitem{Kardar87} M.\ Kardar,
Nucl.\ Phys.\ {\bf B 290}, 582 (1987).

\bibitem{KardarNelsonIncommensurate1985}
Kardar, M., Nelson, D. R. (1985). Commensurate-incommensurate transitions with quenched random impurities. Physical review letters, 55(11), 1157.

\bibitem{LeDoussal1} P.\ Calabrese, P. Le Doussal and A.\ Rosso,
Free-energy distribution of the directed polymer at high temperature,
EPL, {\bf 90}, 20002 (2010).

\bibitem{Dotsenko1} V.\ Dotsenko,
Bethe ansatz derivation of the Tracy-Widom distribution for one-dimensional directed polymers,
EPL, {\bf 90},20003 (2010)

\bibitem{Dotsenko2}  V.\ Dotsenko, Replica Bethe ansatz derivation of the Tracy-Widom distribution of the free energy fluctuations in one-dimensional directed polymers,
J.\ Stat.\ Mech. P07010 (2010).

\bibitem{Kormos}
P. Calabrese, M. Kormos and P. Le Doussal, 
EPL 107 10011, (2014).

\bibitem{Lieb-Liniger} E.H.\ Lieb and W.\ Liniger,
Phys.\ Rev.\  {\bf 130}, 1605 (1963)

\bibitem{McGuire} J.B.\ McGuire,
J.\ Math.\ Phys.\ {\bf 5}, 622 (1964).

\bibitem{Yang} C.N.\ Yang,
Phys.\ Rev.\  {\bf 168}, 1920 (1968)    

\bibitem{gaudin} M.\ Gaudin, 
{\it La fonction d'onde de Bethe}, Paris, Masson, (1983)    

\bibitem{HeckmanOpdam1997} 
Heckman, G. J., Opdam, E. M. Yang's system of particles and Hecke algebras. 
Annals of mathematics, 145(1), 139-173 (1997). 

\bibitem{Brunet-Derrida} E.\ Brunet and B.\ Derrida,
Phys.\ Rev. E {\bf 61}, 6789 (2000)


\bibitem{LargeDevPLD2016}
Le Doussal, P., Majumdar, S. N., Schehr, G. (2016). Large deviations for the height in 1D Kardar-Parisi-Zhang growth at late times. Europhysics Letters, 113(6), 60004.


\bibitem{KPZ-TW1c} T.\ Sasamoto and H.\ Spohn,
J.\ Stat.\ Phys. {\bf 140}, 209 (2010)




\bibitem{ProlhacSpohnPDF2011}
Prolhac, S., Spohn, H. (2011). Height distribution of the Kardar-Parisi-Zhang equation with sharp-wedge initial condition: Numerical evaluations. Physical Review E-Statistical, Nonlinear, and Soft Matter Physics, 84(1), 011119.

\bibitem{DeanFiniteT} 
D.S. Dean, P. Le Doussal, S. N. Majumdar, G. Schehr. 
Phys. Rev. Lett. 114, no. 11 (2015): 110402.



\bibitem{Meerson2016} 
B. Meerson, E. Katzav, A. Vilenkin, Large Deviations of Surface Height in the KardarParisi-Zhang Equation, Physical Review Letters 116, 070601, (2016).

\bibitem{Kraj} For a review see, 
A. Krajenbrink, 
Beyond the typical fluctuations: a journey to the large deviations in the Kardar-Parisi-Zhang growth model. 
PhD thesis, PSL Research University, 2019.
\url{https://theses.hal.science/tel-02537219/document}.

\bibitem{SasorovLargeTime} 
Sasorov, P., Meerson, B., Prolhac, S. (2017). Large deviations of surface height in the 1+ 1-dimensional Kardar-Parisi-Zhang equation: exact long-time results for ?H< 0. Journal of Statistical Mechanics: Theory and Experiment, 2017(6), 063203.

\bibitem{CorwinPLD} 
Corwin, I., Ghosal, P., Krajenbrink, A., Le Doussal, P., Tsai, L. C. (2018). Coulomb-gas electrostatics controls large fluctuations of the Kardar-Parisi-Zhang equation. Physical review letters, 121(6), 060201.


\bibitem{PLD-Flat1} 
P. Calabrese, P. Le Doussal, 
Physical Review Letters 106, 250603, (2011)

\bibitem{PLD-Flat2} 
P. Le Doussal, P. Calabrese, 
J. Stat. Mech. P06001, (2012).



\bibitem{ImamuraStat}
T. Imamura, T. Sasamoto, 
Phys. Rev. Lett. 108, 190603 (2012); 
and J. Stat. Phys. 150, 908-939 (2013).

\bibitem{BBP} 
J. Baik, G. Ben Arous, and S. P\'ech\'e, 
Ann. Probab. 33, 1643 (2005). 

\bibitem{TWASEP1} 
C.A. Tracy and H. Widom, Asymptotics in ASEP with step initial condition,
Comm. Math. Phys. 290, 129-154 (2009).



\bibitem{KPZ-TW1a} T.\ Sasamoto and H.\ Spohn,
Phys.\ Rev.\ Lett.\ {\bf 104}, 230602 (2010)

\bibitem{KPZ-TW1b} T.Sasamoto and H.Spohn,
Nucl.\ Phys.\ {\bf B834}, 523 (2010)


\bibitem{AmirCorwinQuastel}  G.\ Amir, I.\ Corwin and J.\ Quastel,
Comm.\ Pure Appl.\ Math.\ {\bf 64}, 466 (2011)




\bibitem{Schutz1997} 
 SCHUTZ, G. M. (1997). Exact solution of the master equation for the asymmetric exclusion process. J. Stat.
Phys. 88 427-445. 
MR1468391 \url{https://doi.org/10.1007/BF02508478}



\bibitem{MacDo}
A. Borodin and I. Corwin, Macdonald processes, 
Probab. Theory Rel. Fields 158, 225 (2014), arXiv:1111.4408.

\bibitem{CorwinReview}
I. Corwin, 
arXiv:1403.6877.


\bibitem{Veto}
A. Borodin, I. Corwin, P. L. Ferrari. B. Veto, 
Math. Phys. Anal. Geom. 18, 20, (2015).

\bibitem{QuastelFlatASEP2015} 
J. Ortmann, J. Quastel, D. Remenik, 
A Pfaffian representation for flat ASEP
arXiv:1501.05626 
Comm. Pure Appl. Math. 70, 3-89 (2017).

\bibitem{BaxterBook} 
Baxter, Rodney J. Exactly solved models in statistical mechanics. Elsevier, 2016.

\bibitem{FaddeevHouches} 
Faddeev, L.D.: How algebraic Bethe Ansatz works for integrable model. In: Les-Houches
Lecture Notes (1996). arXiv:1407.3367 [math.PR]
Faddeev, L. D. (1996). How algebraic Bethe ansatz works for integrable model. arXiv preprint hep-th/9605187.

\bibitem{Reshetikhin}, N.Reshetikhin: Lectures on the integrability of the 6-vertex model. In: Les-Houches Lecture
Notes (2008). arXiv:1010.5031.

\bibitem{BCG6Vertex2014}
Borodin, A., Corwin, I., Gorin, V.: Stochastic six-vertex model (2014). arXiv:1407.6729.

\bibitem{CorwinPetrovHigherSpinVertex2016} 
Corwin, I., Petrov, L. (2016). Stochastic higher spin vertex models on the line. Communications in Mathematical Physics, 343, 651-700.


\bibitem{BorodinWheeler2018} 
A. Borodin and M. Wheeler. Colored stochastic vertex models and their spectral theory. Ast\'erisque,
(437):ix+225, 2022.
Borodin, A., Wheeler, M. (2018). Coloured stochastic vertex models and their spectral theory. arXiv preprint arXiv:1808.01866.

\bibitem{QuastelKPZFixedPoint} 
K. Matetski, J. Quastel, D. Remenik, 
Acta Math. 227, 115-203 (2021). 



\bibitem{QuastelKP} 
Quastel, J., Remenik, D. (2022, January). KP governs random growth off a 1-dimensional substrate,
arXiv:1908.10353,
In Forum of Mathematics, Pi (Vol. 10, p. e10). Cambridge University Press.

\bibitem{QuastelKPZFPReplica} 
arXiv:1103.3422 
Renormalization fixed point of the KPZ universality class
I. Corwin, J. Quastel, D. Remenik, J. Stat. Phys. 160, 815-834 (2015).


\bibitem{Dotsenko2time}
V. Dotsenko, J. Stat. Mech. P06017 (2013), 
J. Stat. Mech. P06017 (2013), 
[J. Phys. A: Math. Theor. 49 27 (2016), 

\bibitem{deNardisPLD2time}
J. De Nardis, P. Le Doussal, J. Stat. Mech. (2017) 053212, 
and J. Stat. Mech. (2018) 093203. 

\bibitem{deNardisPLDTakeuchi2time}
J. De Nardis, P. Le Doussal, K. A. Takeuchi, Phys. Rev. Lett. 118, 125701 (2017), 


\bibitem{PLDAiry}
P. Le Doussal, 
Phys. Rev. E 96, 060101 (2017).

\bibitem{HH}
T. Halpin Healy, private communication, and talk at College de France
\url{https://youtu.be/sZbavBmtasw} 1h23 mark.

\bibitem{FS2timeStat}
P. L. Ferrari, H. Spohn, SIGMA 12 (2016) 074, arXiv:1602.00486.


\bibitem{Johansson2times} 
K. Johansson, 
Probability Theory and Related Fields 175, no. 3 (2019): 849-895.

\bibitem{BaikMultiTime} 
Baik, J., Liu, Z. (2019). Multipoint distribution of periodic TASEP. Journal of the American Mathematical Society, 32(3), 609-674.

\bibitem{ViragDirectedLandscape2022} 
Dauvergne, D., Ortmann, J., Virag, B. (2022). The directed landscape. Acta Mathematica, 229(2), 201-285.

\bibitem{Virag2020}
Virag, B. (2020). The heat and the landscape I. arXiv preprint arXiv:2008.07241.

\bibitem{KPZeqDirectedWu2023} 
Wu, X. (2023). The KPZ equation and the directed landscape. arXiv preprint arXiv:2301.00547.



\bibitem{HairerRough}
M. Hairer, Rough stochastic PDEs, Commun. on Pure and Applied
Math. 64, 1547-1585 (2011)

\bibitem{HairerKPZ} 
M. Hairer, Solving the KPZ equation, Ann. of Math. 178, 559-664
(2013)

\bibitem{HairerRegularity} 
M. Hairer, A theory of regularity structures, 236pp., to appear in Inventiones mathematicae

\bibitem{IMU}
International Mathematical Union (Seoul 2014)
quoted in \url{https://fiteoweb.unige.ch/~eckmannj/ps_files/HairerSPG.pdf} 

\bibitem{HairerQuastel} 
Hairer, M., Quastel, J. (2018, January). A class of growth models rescaling to KPZ. 
In Forum of Mathematics, Pi (Vol. 6, p. e3). Cambridge University Press.




%
%

%




\bibitem{MedinaKardar}
E. Medina and M. Kardar, Phys. Rev. B 46, 9984 (1992).

\bibitem{Ortuno}
J. Prior, A. M. Somoza and M. Ortuno, Phys. Rev. B 72,
024206 (2005), A. M. Somoza, M. Ortuno and J. Prior, Phys. Rev. Lett.
99, 116602 (2007). 

 \bibitem{Lemarie} 
Mu, S., Gong, J., Lemari\'e, G. (2024). Kardar-Parisi-Zhang Physics in the Density Fluctuations of Localized Two-Dimensional Wave Packets. Physical Review Letters, 132(4), 046301.

\bibitem{SomozaPLD} 
Somoza, A. M., Le Doussal, P., Ortuno, M. (2015). Unbinding transition in semi-infinite two-dimensional localized systems. Physical Review B, 91(15), 155413.

\bibitem{ViragGraphene}
 M. Kotowski and B. Virag, Tracy-Widom fluctuations in 2D random Schrodinger operators, Comm. Math.
Phys. 370 (2019), no. 3, 873-893.

\bibitem{BarraquandVirag} 
Barraquand, G., Corwin, I., Dimitrov, E. (2021). Maximal free energy of the log-gamma polymer. arXiv preprint arXiv:2105.05283.

\bibitem{KesslerLevine1991} 
D. A. Kessler, H. Levine, and Y. Tu, Phys. Rev. A 43, 4551
(1991).

\bibitem{KardarLines} 
 M. Kardar, Nonequilibrium dynamics of interfaces and lines,
Phys. Rep. 301 (1998) 85112.

\bibitem{MovingGlass} 
[P. Le Doussal and T. Giamarchi, Moving glass theory of driven
lattices with disorder, Phys. Rev. B 57 (1998) 1135611403.

\bibitem{ParisiQKPZ} 
Parisi, G. (1992). On surface growth in random media. Europhysics Letters, 17(8), 673.




\bibitem{Anisotropic} 
L.-H. Tang, M. Kardar and D. Dhar, Driven depinning in
anisotropic media, Phys. Rev. Lett. 74 (1995) 920-3.


\bibitem{DirectedPerco} 
L.-H. Tang and H. Leschhorn, Pinning by directed percolation,
Phys. Rev. A 45 (1992) R8309-12.

\bibitem{Atis} 
Atis, S., Dubey, A. K., Salin, D., Talon, L., Le Doussal, P., Wiese, K. J. (2015). Experimental evidence for three universality classes for reaction fronts in disordered flows. Physical review letters, 114(23), 234502.

\bibitem{ErtasKardar1992} 
Ertas, D., Kardar, M. (1992). Dynamic roughening of directed lines. Physical review letters, 69(6), 929.

\bibitem{Hwa} 
Hwa, T. (1992). Nonequilibrium dynamics of driven line liquids. Physical review letters, 69(10), 1552.

\bibitem{Ramaswamy1997} 
 R. Lahiri and S. Ramaswamy, Phys. Rev. Lett. 79, 1150
(1997).

\bibitem{CoupledKPZSpohn2013} 
Ferrari, P. L., Sasamoto, T., Spohn, H. (2013). Coupled Kardar-Parisi-Zhang equations in one dimension. Journal of Statistical Physics, 153, 377-399.

\bibitem{SpohnHydroReview}
Herbert Spohn. Nonlinear fluctuating hydrodynamics for anharmonic chains. Journal of Statistical
Physics, 154:1191, 2014

\bibitem{SpohnStoltz2015} 
Spohn, H., Stoltz, G. (2015). Nonlinear fluctuating hydrodynamics in one dimension: the case of two conserved fields. Journal of Statistical Physics, 160, 861-884.

\bibitem{PopkovSchutz2015} 
V. Popkov, A. Schadschneider, J. Schmidt, and G.M. Schutz, Fibonacci family of dynamical universality classes, Proc.
Natl. Acad. Science (USA) 112 12645-12650 (2015).

\bibitem{PopkovModeCoupling2016} 
V. Popkov, A. Schadschneider, J. Schmidt, G.M. Schutz, Exact scaling solution of the mode coupling equations for nonlinear fluctuating hydrodynamics in one dimension, J. Stat. Mech. 093211 (2016).

\bibitem{PopkovSchutzGolden2024} 
V. Popkov, G. M. Schutz, Quest for the golden ratio universality class, 
arXiv:2310.19116, Physical Review E 109, 044111 (2024).

\bibitem{Delacretaz} 
Delacretaz, L. V., Fitzpatrick, A. L., Katz, E., Walters, M. (2022). Thermalization and hydrodynamics of two-dimensional quantum field theories. SciPost Physics, 12(4), 119.


\bibitem{Prosen2017} 
M. Ljubotina, M. Znidaric, and T. Prosen, Spin diffusion
from an inhomogeneous quench in an integrable system,
Nat. Commun. 8, 16117 (2017).


\bibitem{Prosen2019} 
Marko Ljubotina, Marko Znidaric, and Tomaz Prosen. Kardar-Parisi-Zhang physics in the quantum
Heisenberg magnet. Physical Review Letters, 122:210602, 2019.

\bibitem{VasseurScaling2019} 
S. Gopalakrishnan and R. Vasseur, Kinetic theory of spin
diffusion and superdiffusion in XXZ spin chains, Phys.
Rev. Lett. 122, 127202 (2019).

\bibitem{deNardisBurgers} 
J. De Nardis, M. Medenjak, C. Karrasch, and E. Ilievski
Anomalous spin diffusion in one-dimensional antiferromagnets, Phys. Rev. Lett. 123, 186601 (2019).


\bibitem{MooreNature2021} 
A. Scheie, N. E. Sherman, M. Dupont, S. E. Nagler, M. B. Stone, G. E. Granroth, J. E. Moore,
and D. A. Tennant. Detection of KardarParisiZhang hydrodynamics in a quantum Heisenberg
spin-1/2 chain. Nature Physics, 17:726, 2021.

\bibitem{Bloch2022} 
David Wei, Antonio Rubio-Abadal, Bingtian Ye, Francisco Machado, Jack Kemp, Kritsana Srakaew,
Simon Hollerith, Jun Rui, Sarang Gopalakrishnan, Norman Y. Yao, Immanuel Bloch, and
Johannes Zeiher. Quantum gas microscopy of KardarParisiZhang superdiffusion. Science,
376:716, 2022.

\bibitem{Simulations}  
 B. Ye, F. Machado, J. Kemp, R. B. Hutson, and N. Y.
Yao, Universal Kardar-Parisi-Zhang dynamics in integrable quantum systems, Phys. Rev. Lett. 129, 230602
(2022),
A. Das, M. Kulkarni, H. Spohn, and A. Dhar, KardarParisi-Zhang scaling for an integrable lattice LandauLifshitz spin chain, Phys. Rev. E 100, 042116 (2019).

\bibitem{Universal} 
V. B. Bulchandani, S. Gopalakrishnan, and E. Ilievski,
Superdiffusion in spin chains, J. Stat. Mech. 2021,
084001 (2021), S. Gopalakrishnan and R. Vasseur, Superdiffusion from
nonabelian symmetries in nearly integrable systems,
Annu. Rev. Condens. Matter Phys. 15, 159 (2024).


\bibitem{Bulchandani2020} 
Bulchandani, V. B. (2020). Kardar-Parisi-Zhang universality from soft gauge modes. Physical Review B, 101(4), 041411.

\bibitem{JacopoCoupled2023} 
Jacopo De Nardis, Sarang Gopalakrishnan, and Romain Vasseur. Nonlinear fluctuating
hydrodynamics for Kardar-Parisi-Zhang scaling in isotropic spin chains. Physical Review Letters,
131:197102, 2023.

\bibitem{SpohnDegenerate2024}
Universality in coupled stochastic Burgers systems with degenerate flux Jacobian
arXiv:2401.06399,
Dipankar Roy, Abhishek Dhar, Konstantin Khanin, Manas Kulkarni, Herbert Spohn. 

\bibitem{Takeuchi2024}
Kazumasa A. Takeuchi, Jacopo De Nardis, Ofer Busani, Patrik L. Ferrari, Romain Vasseur,
arXiv:2406.07150,
Partial yet definite emergence of the Kardar-Parisi-Zhang class in isotropic spin chains



\bibitem{Nahum2017} 
Nahum, A., Ruhman, J., Vijay, S., Haah, J. (2017). Quantum entanglement growth under random unitary dynamics. Physical Review X, 7(3), 031016.

\bibitem{Meakin1986} 
P. Meakin, P. Ramanlal, L. M. Sander, and R. C. Ball,
Ballistic Deposition on Surfaces, Phys. Rev. A 34, 5091
(1986).

\bibitem{Nahum2019} 
Zhou, T., Nahum, A. (2019). Emergent statistical mechanics of entanglement in random unitary circuits. Physical Review B, 99(17), 174205.

\bibitem{Nahum2018} 
Nahum, Adam, Sagar Vijay, and Jeongwan Haah. "Operator spreading in random unitary circuits." Physical Review X 8.2 (2018): 021014.

\bibitem{BernardPLD2020} 
Bernard, D., Le Doussal, P. (2020). Entanglement entropy growth in stochastic conformal field theory and the KPZ class. Europhysics Letters, 131(1), 10007.

\bibitem{Grinstein1993} 
G. Grinstein, D. Mukamel, R. Seidin, and C. H. Bennett,
Temporally Periodic Phases and Kinetic Roughening, Phys.
Rev. Lett. 70, 3607 (1993).

\bibitem{AltmanToner2015} 
Altman, E., Sieberer, L. M., Chen, L., Diehl, S. Toner, J. Two-dimensional superfluidity of
exciton polaritons requires strong anisotropy. Phys. Rev. X 5, 011017 (2015).

\bibitem{DiehlSuperfluid2015} 
He, L., Sieberer, L. M., Altman, E., Diehl, S. (2015). Scaling properties of one-dimensional driven-dissipative condensates. Physical Review B, 92(15), 155307.



\bibitem{DiehlKPZVortex2017} 
He, L., Sieberer, L. M. Diehl, S. Space-time vortex driven crossover and vortex
turbulence phase transition in one-dimensional driven open condensates. Phys. Rev. Lett.
118, 085301 (2017).

\bibitem{PolaritonNature2022} 
Fontaine, Q., Squizzato, D., Baboux, F., Amelio, I., Lemaitre, A., Morassi, M., ... Bloch, J. (2022). KardarParisiZhang universality in a one-dimensional polariton condensate. Nature, 608(7924), 687-691.



%
%
%
%
%
%
%
%


\bibitem{CuernoSynchronization2024}
R Gutierrez, R Cuerno, Kardar-Parisi-Zhang universality class in the synchronization of oscillator lattices with time-dependent noise, arXiv:2407.15634. 

\bibitem{TakeuchiChaos2021} 
For review see, Fukai, Y. T., Takeuchi, K. A. (2021). Initial perturbation matters: Implications of geometry-dependent universal KardarParisiZhang statistics for spatiotemporal chaos. Chaos: An Interdisciplinary Journal of Nonlinear Science, 31(11).

\bibitem{RWRECorwinBarraquand} 
Barraquand, G., Corwin, I. (2017). Random-walk in beta-distributed random environment. Probability Theory and Related Fields, 167(3), 1057-1116.

\bibitem{TTPLD2017} 
Le Doussal, P., Thiery, T. (2017). Diffusion in time-dependent random media and the Kardar-Parisi-Zhang equation. Physical Review E, 96(1), 010102.

\bibitem{NelsonRangeExpansion2007}
Hallatschek, O., Hersen, P., Ramanathan, S., Nelson, D. R. (2007). Genetic drift at expanding frontiers promotes gene segregation. Proceedings of the National Academy of Sciences, 104(50), 19926-19930.







   


	
	
%
%
%
%
%
%
%

%
%
%
%


	





%
%
%
%








%
%
%
	
	



%
%
%
%
%
%
%
%
%
%
%
%
%
%
%
%
%
%
%
%
%
%
%
%
%
%
%
%
%
%
%
%
%
%
%
%
%
%
%
%
%
%
%
%
%
%
%
%
%
%
%
%
%
%
%
%
%
%
%
%
%
%
%
%
%
%
%
%
%
%
%
%
%
%
%
%
%
%
%
%
%
%
%
%
%
%
%
%
%
%
%
%
%
%
%
%
%
%
%
%



		
\end{thebibliography}
\end{document}